\documentclass[aps,prd,onecolumn,showpacs,showkeys,preprintnumbers,amsmath,amssymb,a4paper]{revtex4-1}
\usepackage{amssymb}
\usepackage{graphics}
\usepackage{psfrag}
\usepackage{epsfig}
\def\bea{\begin{eqnarray}}
\def\eea{\end{eqnarray}}
\def\be{\begin{equation}}
\def\ee{\end{equation}}

\newcommand{\pbarp}{{\bar p p}}
\newcommand{\nbarn}{{\bar n  n}}
\newcommand{\NbarN}{{\bar N  N}}
\newcommand{\ebare}{{e^+e^-}}

\newcommand{\Si}{{\Sigma}}
\newcommand{\La}{{\Lambda}}
\newcommand{\ybary}{{\bar YY}}
\newcommand{\lbarl}{{\bar \Lambda \Lambda}}

\newcommand{\lbars}{{\bar \Lambda \Sigma}}
\newcommand{\sbarl}{{\bar \Sigma^0 \Lambda}}
\newcommand{\xbarx}{{\bar \Xi \Xi }}

\newcommand{\sbars}{{\bar \Sigma \Sigma}}
 \newcommand{\spbarsp}{{\bar \Sigma^- \Sigma^+}}
 \newcommand{\sobarso}{{\bar \Sigma^0 \Sigma^0}}
 \newcommand{\smbarsm}{{\bar \Sigma^+ \Sigma^-}}
\newcommand{\xmbarxm}{{\bar \Xi^+ \Xi^-}}
\newcommand{\xobarxo}{{\bar \Xi^0 \Xi^0}}
\begin{document}
 \title{Hyperon electromagnetic form factors in the timelike region}
%
\author{Johann Haidenbauer$^1$, Ulf-G. Mei{\ss}ner$^{2,1,3}$, Ling-Yun Dai$^4$}
\affiliation{
$^1${Institute for Advanced Simulation, Institut f{\"u}r Kernphysik and
J\"ulich Center for Hadron Physics, Forschungszentrum J{\"u}lich, D-52425 J{\"u}lich, Germany} \\
$^2${Helmholtz Institut f\"ur Strahlen- und Kernphysik and Bethe Center
    for Theoretical Physics, Universit\"at Bonn, D-53115 Bonn, Germany} \\
$^3${Tbilisi State University, 0186 Tbilisi, Georgia}  \\
$^4${School of Physics and Electronics, Hunan University, Changsha 410082, China}
}

\begin{abstract}
Electromagnetic form factors of hyperons ($\Lambda$, $\Sigma$, $\Xi$) in the 
timelike region, accessible in the reaction $e^+e^- \to \bar YY$, are studied. 
The focus is on energies close to the reaction thresholds, where 
the properties of these form factors are significantly influenced by the  
interaction in the final $\bar YY$ system. This interaction is taken into 
account in the calculation, utilizing $\bar YY$ potential models that have 
been constructed by the J\"ulich group for the analysis of data from 
the reaction $\bar pp \to \bar YY$ in the past. 
The enhancement of the effective form factor for energies close to the 
threshold, seen in experiments of $e^+e^- \to \bar \Lambda \Lambda$ 
and $e^+e^- \to \bar \Sigma^0\Lambda$, is reproduced. 
With regard to the reactions $e^+e^- \to \bar \Sigma^- \Sigma^+, \ 
\bar\Sigma^0\Sigma^0, \ \bar\Sigma^+\Sigma^-$
a delicate interplay between the three channels is observed in 
the results at low energies, caused by the $\bar\Sigma\Sigma$ interaction. 
Predictions for the electromagnetic form factors $G_M$ and $G_E$ in the timelike 
region are presented for the $\Lambda$, $\Sigma$, and $\Xi$ hyperons. 
\end{abstract}

\pacs{13.40.Gp, 13.66.Bc, 12.39.Pn}

\maketitle

\section{Introduction}
\label{sec:intro}

Our knowledge on the electromagnetic form factors (EMFFs) of the nucleon
in the timelike region has significantly improved over the last years.
Information on these quantities is accessible in the $\pbarp$ annihilation
process $\pbarp \to \ebare$ \cite{Bardin:1994,Singh:2016}, 
and likewise in the reactions $\ebare \to \pbarp$ and $\ebare \to \nbarn$ 
from where most of the recent data emerged 
\cite{Aubert:2006,Lees:2013,Achasov:2014,Ablikim:2015,Akhmetshin:2015} and 
where experimental efforts continue 
\cite{Akhmetshin:2019,Ablikim:2020P}.
For a general overview, see the review \cite{Denig:2013}.
Regarding the EMFFs of hyperons ($\Lambda$, $\Sigma$, $\Xi$, ...) in the timelike region, 
the development is somewhat lagging behind. Nonetheless, while only a handful 
of experiments \cite{Bisello:1990,Aubert:2007} and 
theoretical studies \cite{Baldini:2007,Dalkarov:2009,Faldt:2013} 
were available a few years ago, the situation has changed definitely 
in recent times with various experimental
\cite{Dobbs:2014,Dobbs:2017,Niiyama:2018,Ablikim:2017,Ablikim:2019,Ablikim:2020,Ablikim:2020X1,Ablikim:2020X2}   
and theoretical
\cite{Haidenbauer:2016,Cao:2018,Yang:2018,Yang:2019,Xiao:2019,Dubnicka:2018,Faldt:2017,Perotti:2019,Ferroli:2020,Ramalho:2020}
activities.
To some extent this is surprising given that the study of hyperons offers 
advantages in accessing more detailed information on the EMFFs.
Specifically, the self-analyzing character of the weak decay of the
hyperons can be exploited to determine also the polarization as well 
as spin-correlation parameters for reaction such as $\ebare \to \lbarl$,
$\ebare \to \sbars$, or even $\ebare \to \xbarx$.
Those observables allow one to determine not only bulk properties such as 
the reaction cross section or the (effective) form factor but also the relative 
magnitude of the two EMFFs $G_E$ and $G_M$ and even the relative phase 
between them \cite{Denig:2013,Faldt:2016}. 
A first measurement of that kind has been reported recently by the BESIII collaboration
for the reaction $\ebare \to \lbarl$ at $\sqrt{s}= 2.396$~GeV \cite{Ablikim:2019}. 

In the present work we take that achievement as motivation to explore the 
electromagnetic form factors of other hyperons in the timelike region. 
A study at their properties is timely given that cross sections for the 
reactions $\ebare \to \sbarl$ \cite{Aubert:2007}, $\ebare \to \sbars$ 
\cite{Aubert:2007,Dobbs:2017,Ablikim:2020}, and $\ebare \to \xbarx$ 
\cite{Ablikim:2020X1,Ablikim:2020X2}
have become already available, although not yet in a quantity and quality 
comparable to $\ebare \to \lbarl$ \cite{Aubert:2007,Ablikim:2017,Ablikim:2019}.
Measuring spin-dependent observables and, thereby, determining the properties
of the form factors $G_M$ and $G_E$ is the logical next step   
\cite{Ablikim:2020R}.
We study the reactions $\ebare \to \ybary$ near their threshold where the
essential input is the $\ybary$ interaction in the final state. In particular, 
we provide predictions for the energy dependence of the reaction cross sections
and also for the magnitude and relative phase of the form factors $G_M$ and $G_E$ .
The calculation is done in close analogy to our analysis of the reaction
$\ebare \to \pbarp$ \cite{Haidenbauer:2014} where the interaction in the 
$\pbarp$ system has been taken into account rigorously. It turned out 
that, after including the $\pbarp$ final-state interaction (FSI),
available data on the proton form factor but also differential 
cross sections and the ratio $|G_E/G_M|$ could be reproduced fairly well.
The employed $\NbarN$ interaction had been derived within chiral effective 
field theory \cite{Kang:2013,Dai:2017} fitted to results of a $\pbarp$ 
partial-wave analysis \cite{Zhou:2012}. 
We note in passing that this approach has been also successfully applied
to the reaction 
$\ebare \to \bar \Lambda_c \Lambda_c$ \cite{Pakhlova:2008,Ablikim:2017Lc,Dai:2017Lc}.

Unfortunately, with regard to the $\ybary$ interaction we are not in such a 
comfortable situation as for $\NbarN$. Without the option for obtaining direct empirical 
information about the $\ybary$ force, constraints for that interaction can be only 
inferred from studies of FSI effects in reactions such as $\pbarp \to \ybary$ 
and/or the analysis of $\ybary$ correlation functions measured in heavy-ion 
collisions \cite{Acharya:2020}. 
By far the best investigated case is $\lbarl$, where the reaction $\pbarp\to\lbarl$ 
has been extensively measured in the PS185 experiment at LEAR and where data 
are available (from the reaction threshold up to about $2.4$~GeV)
for total and differential cross-sections, but also for spin-dependent
observables \cite{PS1851,PS1852,PS1853,Barnes:2000,PS185},  
thanks to the aforementioned self-analyzing weak $\Lambda$ decay. 
The level of information is much less satisfactory for other channels like 
$\pbarp \to \sbarl$ and/or $\pbarp \to \sbars$ \cite{Barnes:1997,Johansson:1999,TordSS}
and, in particular, for $\pbarp \to \xbarx$. 
 
In the present investigation we employ a phenomenological $\ybary$ model 
that has been developed by the J\"ulich group for the analysis of those 
PS185 data in the past
\cite{Haidenbauer:1991,Haidenbauer:1992A,Haidenbauer:1992,Haidenbauer:1993,Haidenbauer:1993X}.
In case of the $\lbarl$ interaction there are several variants which all describe the
$\pbarp\to\lbarl$ data quite well \cite{Haidenbauer:1991,Haidenbauer:1992A,Haidenbauer:1992}.
Indeed, these variants have been already used by us for calculating the 
$\ebare \to \lbarl$ reaction \cite{Haidenbauer:2016} and yielded cross sections 
well in line with the data. Interestingly, it turned out that one of the interactions even 
reproduces roughly the BESIII results on the ratio and phase of the EMFFs \cite{Ablikim:2019}. 
The interactions utilized for the $\sbarl$ and $\sbars$ channels were also tuned to 
the pertinent experimental information from PS185 \cite{Haidenbauer:1993}. 
However, due to the limited number of data, these interactions are certainly not
well constrained. Nonetheless, we believe that it is instructive to explore the 
corresponding predictions at the present stage. They should be considered 
mainly as a guideline and as a stimulation for future experiments. 
Of course, ultimately (complete) measurements of the reactions $\ebare \to \sbarl$ 
and $\ebare \to \sbars$ will provide new additional and rather stringent constraints 
on the $\sbarl$ and $\sbars$ forces and, in turn, should be used to pin down the 
corresponding interactions more reliably in the future.

The paper is organized as follows: 
In Sect.~\ref{sec:form} we describe the employed formalism and in Sect.~\ref{sec:model}
we summarize the properties of the $\ybary$ potential models used in the
calculations.
Numerical results for the $\ebare \to \ybary$ reactions are presented in
Sect.~\ref{sec:results}. Specifically, we show the outcome for the cross sections 
and for the effective form factors, and we provide predictions for the properties of the
electromagnetic form factors $G_M$ and $G_E$. 
The paper closes with a Summary. 

\section{Formalism}
\label{sec:form}

The reaction $\ebare \to \ybary$ is studied with the same formalism as
applied by us to $\ebare \to \pbarp$, which was developed and described 
in detail in Ref.~\cite{Haidenbauer:2014}. We summarize it briefly below. 
We adopt the standard conventions \cite{Denig:2013} so that the differential 
cross section for the reactions $\ebare \to \ybary$ is given by 
\be
\frac{d\sigma}{d\Omega} = \frac{\alpha^2\beta}{4 s}
~C(s)~
\left[\left| G^Y_M(s) \right|^2 (1+{\rm cos}^2\theta) +
\frac{4M_{Y}^2}{s} \left| G^Y_E(s) \right|^2 {\rm sin}^2\theta \right]~{\rm .}
\label{eq:diff}
\ee
where $Y$ generically denotes the hyperons $\La$, $\Si$, and $\Xi$,
and $\bar Y$ stands for the corresponding anti-particles.
Here, $\alpha = 1/137.036$ is the  fine-structure constant and
$\beta=k_{Y}/k_e$ a phase-space factor, where $k_{Y}$ and $k_e$ are the
center-of-mass three-momenta in the $\ybary$ and $\ebare$ systems, respectively,
related to the total energy via $\sqrt{s} = 2\sqrt{M_{Y}^2+k_p^2} = 2\sqrt{m_e^2+k_e^2}$.
Further, $m_e \, (M_{Y})$ is the electron ($Y$) mass. 
The $S$-wave Sommerfeld-Gamow factor $C(s)$ is given by $C = y/(1-e^{-y})$
with $y = \pi \alpha M_{Y} /k_{Y}$. Of course, for uncharged hyperons 
($\La$, $\Si^0$, $\Xi^0$) we have $C(s)\equiv 1$. 
$G_E$ and $G_M$ are the electric and magnetic form factors of the hyperons in question. 
In general, we omit the superscript $Y$ from the $G$'s in the following because 
it is anyhow clear from the context which hyperon is discussed.
The cross section as written in Eq.~({\ref{eq:diff}) results from the one-photon
exchange approximation and by setting the electron mass $m_e$ to zero
(in that case $\beta = 2k_{Y}/\sqrt{s}$). We will restrict
ourselves throughout this work to the one-photon exchange so that the total
angular momentum is fixed to $J=1$ and the $\ebare$ and $\ybary$ systems can be only in
the partial waves $^3S_1$ and $^3D_1$. We use the standard spectral notation
$^{(2S+1)}L_J$, where $S$ is the total spin and $L$ the orbital angular momentum.

The integrated reaction cross section is readily found to be
\be
\sigma_{\ebare \to \ybary} = \frac{4 \pi \alpha^2 \beta}{3s}~C(s)~
\left [ \left| G_M(s) \right|^2 + \frac{2M_{Y}^2}{s} \left| G_E(s) \right|^2 \right ]~{\rm .}
\label{eq:tot}
\ee

Another quantity used in various analyses is the effective baryon form factor $G_{\rm eff}$
which is defined by
\be
|G_{\rm eff} (s)|=\sqrt{\sigma_{\ebare\rightarrow \ybary} (s)\over {4\pi\alpha^2
\beta \over 3s} ~C(s)   
\left [1 +\frac{2M^2_{Y}}{s}\right ]} \ .
\label{eq:Geff}
\ee

In order to implement the FSI we perform a partial wave projection of the
$\ebare \to \ybary$ amplitudes. The corresponding formalism is documented
in various publications in the literature. We follow here the procedure described
in detail in the Appendices~B and C of Ref.~\cite{Holzenkamp89}.
Then we end up with four amplitudes, corresponding to the
coupling between the $\ebare$ and the $\ybary$ systems and
the coupled $^3S_1$-$^3D_1$ partial waves.
We can write these in the form $F_{L\,L'}$, where $L' (L)=0,2$
characterizes the orbital angular momentum in the initial (final)
state. The explicit expressions for the reaction $\ebare \to \ybary$
are
\bea
\nonumber
F^{\mu,\nu}_{2\,2} &=& -\frac{2\alpha}{9} \left[ G_M - \frac{2 M_Y}{\sqrt{s}} G_E\right]
\left[ 1 - \frac{2 m_e}{\sqrt{s}} \right]~, \\
\nonumber
F^{\mu,\nu}_{0\,0} &=& -\frac{4\alpha}{9} \left[ G_M + \frac{M_Y}{\sqrt{s}} G_E\right]
\left[ 1 + \frac{m_e}{\sqrt{s}} \right]~, \\
\nonumber
F^{\mu,\nu}_{0\,2} &=& -\frac{2\sqrt{2}\alpha}{9} \left[ G_M + \frac{M_Y}{\sqrt{s}} G_E\right]
\left[ 1 - \frac{2 m_e}{\sqrt{s}} \right]~, \\
F^{\mu,\nu}_{2\,0} &=& -\frac{2\sqrt{2}\alpha}{9} \left[ G_M - \frac{2 M_Y}{\sqrt{s}} G_E\right]
\left[ 1 + \frac{m_e}{\sqrt{s}} \right] \ .
\label{eq:pwa}
\eea
The superscripts in Eq.~(\ref{eq:pwa}) symbolize the channels where $\nu = \ebare$ and $\mu = \ybary$.

It is obvious from Eq.~(\ref{eq:pwa}) that the amplitude $F^{\mu,\nu}_{L\,L'}$ can
be written as a product of factors. This is simply a consequence of
the one-photon exchange which amounts to an $s$-channel pole diagram in
the reactions $\ebare \to \ybary$. The factors correspond to
the $\ebare\gamma$ and $\ybary\gamma$ vertices, respectively, and
reflect whether the coupling occurs in an $S$ or $D$ wave.
Thus, we can write the amplitude in the form ($L,\,L'=0,2$)
\bea
\nonumber
F^{\mu,\nu}_{L\,L'} &=& -\frac{4\alpha}{9} \ f^{\mu}_L \ f^{\nu}_{L'}, \ \ \  {\rm with}  \\
f^{\mu}_0 &=& \left(G_M + \frac{M_{Y}}{\sqrt{s}} G_E\right), \
f^{\mu}_2= \frac{1}{\sqrt{2}}  \left(G_M - \frac{2 M_{Y}}{\sqrt{s}} G_E\right)~,
\label{FOFA}
\eea
and similar expressions for $f^{\nu}_L$, the vertex functions of the $\ebare$ pair. 
The FSI effects due to the $\ybary$ interaction influence only the
$\ybary$ vertex and that means only $f^{\mu}_L$, see Fig.~1 in Ref.~\cite{Haidenbauer:2014}. 
These effects can be calculated rigorously within our formalism \cite{Haidenbauer:2014} 
after a proper generalization in order to take into account the possible coupling between 
the various $\ybary$ channels. Considering such a coupling, Eq.~(8) in the above reference
then reads
\begin{eqnarray}
&&f^{\mu '}_{L'}(k;E_k)=f^{0\,({\mu '})}_{L'}(k)+
\sum_{\mu} \sum_{L}\int_0^\infty \frac{dpp^2}{(2\pi)^3} \, f^{0\,(\mu)}_{L}(p)
\frac{1}{2E_k-2E^{\mu}_p+i0^+}T^{\mu,\mu '}_{LL'}(p,k;E_k) \ , 
\label{FSI}
\end{eqnarray}
where $\mu,\,\mu '$ stand for any of the $\ybary$ channels. 
For each $\ybary$ channel the first term on the right-hand side, the so-called
Born term, represents the bare $\ybary$ production vertex
$f^{0\,({\mu '})}_L$ and the integral provides the dressing of this vertex
via rescattering in all the (coupled) $\ybary$ channels.
The quantity $T^{\mu,\mu '}_{LL'}(p,p';E_k)$ is the $\ybary$ scattering
amplitude in the coupled $^3S_1$-$^3D_1$ partial wave and is
the solution of a corresponding Lippmann-Schwinger equation:
\begin{eqnarray}
&&T^{\mu '',\mu '}_{L''L'}(p'',p';E_k)=V^{\mu '',\mu '}_{L''L'}(p'',p')+
\sum_{\mu}\sum_{L}\int_0^\infty \frac{dpp^2}{(2\pi)^3} \, 
V^{\mu '',\mu}_{L''L}(p'',p)
\frac{1}{2E_k-2E^{\mu}_p+i0^+}T^{\mu,\mu '}_{LL'}(p,p';E_k)~, 
\label{LS}
\end{eqnarray}
see Ref.~\cite{Haidenbauer:1993}. In the above equations,
$2\,E_k=2\sqrt{M^2_Y+k^2}=\sqrt{s}$, where $k$ is the $\ybary$ on-shell momentum.

The bare $\ybary \gamma$ vertex functions, $f^{0\,(\mu)}_L$ ($L$=0,2) in 
Eq.~(\ref{FSI}), can be written in terms of bare EMFFs, 
$G^{0\,(\mu)}_E$ and $G^{0\,(\mu)}_M$, in complete analogy to Eq.~(\ref{FOFA}).
In general, $f^{0\,(\mu)}_0$ and $f^{0\,(\mu)}_2$ can depend on the total 
energy and on the (off-shell) momentum of the $\ybary$ system. Furthermore, they
are complex because $f^{0\,(\mu)}_L$ (or, equivalently, $G^{0\,(\mu)}_E$ 
and $G^{0\,(\mu)}_M$) involve implicitly contributions from intermediate 
mesonic states such as $\gamma \to \pi^+\pi^- \to \ybary$, 
$\gamma \to K^+ K^- \to \ybary$, etc. 

In the present study we assume that the entire energy dependence of the 
dressed vertex functions $f^{\mu}_L$ is generated by the FSI alone and that
$f^{0\,(\mu)}_0$ and $f^{0\,(\mu)}_2$ themselves are energy-independent. 
In particular, we interpret the explicit
dependence of $f^{0\,(\mu)}_L$ on $\sqrt{s}$ that is implied by Eq.~(\ref{FOFA}) as a
dependence on the momentum of the $\ybary$ system. Accordingly, we use
\bea
f^{0\,(\mu)}_0(p) &=& \left(G^{0\,(\mu)}_M + \frac{M_Y}{2E_p}\, G^{0\,(\mu)}_E\right) =
\left(G^{0\,(\mu)}_M + \frac{M_Y}{2\sqrt{M_Y^2+p^2}}\, G^{0\,(\mu)}_E\right)~,\nonumber \\
f^{0\,(\mu)}_2(p) &=& \frac{1}{\sqrt{2}}\left(G^{0\,(\mu)}_M - \frac{M_Y}{E_p}\,G^{0\,(\mu)}_E\right)
= \frac{1}{\sqrt{2}}\left(G^{0\,(\mu)}_M - \frac{M_Y}{\sqrt{M_Y^2+p^2}}\,G^{0\,(\mu)}_E\right)~,
\label{FOFA0}
\eea
for the bare vertex functions, where $p$ is the center-of-mass momentum
in the $\ybary$ system.
Note that the replacement $\sqrt{s} \to 2E_p$ is required in order to
guarantee the correct threshold behavior of the {$D$-wave} vertex function
$f^0_2(p)$ which has to behave like $\propto p^2$. Indeed, Eqs.~(\ref{FOFA}) 
or (\ref{FOFA0}) reveal that the condition $G^0_E=G^0_M$ and/or 
$G_E=G_M$ in the timelike region at the $\ybary$ thresholds \cite{Denig:2013}
is equivalent to implementing the proper threshold behavior of the $D$-wave amplitude.

We assume that $G^0_E$ and $G^0_M$ are constants 
which automatically implies that we have to set $G^0_E=G^0_M$.
$G^0_E$ ($G^0_M$) were taken to be real in our studies of the reactions 
$\ebare\to \pbarp$ and $\ebare\to \lbarl$ because any overall
phase drops out in the evaluation of observables. In that case
there is only a {\em single} free parameter in the calculation, 
which is basically a normalization constant.
In a coupled-channel approach, cf. Eq.~(\ref{FSI}), there is a $G^0_E$ ($G^0_M$) 
for each channel and only one of them can be chosen to be real. 
However, in the present investigations of the hyperon form factores we 
will consider such a coupling only between the $\sbars$ channels, i.e. 
for $\spbarsp$, $\sobarso$, $\smbarsm$. In this case the reaction thresholds 
are very close together and the channels strongly influence each other.  
Moreover, experimental information is available for all three channels. 
Then there are three constants that need to be fixed from the data,
where one of them can be chosen real. The bare form factors for the
other reactions ($\lbarl$, $\sbarl$, $\xbarx$) are fixed independently
and are all chosen to be real. 
 
The bare vertex functions $f^0_0$ and $f^0_2$ are calculated
from Eq.~(\ref{FOFA0}) and inserted into Eq.~(\ref{FSI}).
Due to the FSI the resulting dressed vertex functions $f_0$ and
$f_2$ are always energy-dependent and complex.
Inverting Eq.~(\ref{FOFA}) we can obtain $G_E$ and $G_M$ and
then evaluate any $\ebare \to \ybary$ observable based on the formulae 
provided at the beginning of this section. 
$G_E$ and $G_M$ are also always complex and, in general, 
$G_E \neq G_M$, where the difference is likewise solely due to the FSI.

Note that the formulae given above apply for $\ebare \to \sbarl$
as well, once the $Y$ mass is replaced by $(M_{\Si^0}+M_{\La})/2$ and, accordingly,
$E_p$ by $\left(\sqrt{M_{\Si^0}^2+p^2}+\sqrt{M_{\La}^2+p^2}\right)/2$. 


\section{The {\boldmath$\ybary$} interaction}
\label{sec:model}

In the present work we utilize a $\ybary$ interaction that has been 
established in studies of antihyperon-hyperon production in $\pbarp$ 
collisions \cite{Haidenbauer:1991,Haidenbauer:1992,Haidenbauer:1993,Haidenbauer:1993X}.
In this model of the J\"ulich group the hyperon-production reaction is 
considered within a coupled-channel approach. This allows one to take 
into account rigorously the effects of the initial ($\pbarp$) and 
final ($\ybary$) state interactions.
The interaction in the various $\ybary$ channels and the transition 
between them is described by meson exchanges whose parameters are
taken over from studies of the $NN$ \cite{obepf} and $YN$ interactions
\cite{Holzenkamp89} on the basis of a $G$-parity transformation. 
The strangeness production process $\pbarp\to \ybary$ is given by the 
exchange of strange mesons~\cite{Haidenbauer:1991,Haidenbauer:1992}.
Annihilaton in the $\ybary$ channels is taken into account by phenomenological 
optical potentials. These potentials involve free parameters, which have been
fixed by a fit to the $\pbarp \to \ybary$ observables. 
We want to emphasize that the experimental constraints in the various $\ybary$ 
channels are of rather different quality. 
The reaction $\pbarp \to \lbarl$ has been extensively investigated in the PS185 
experiment and data are available for total and differential cross sections as
well as for spin-dependent observables \cite{PS1851,PS1852,PS1853,Barnes:2000}, 
down to energies very close to the reaction threshold, see e.g. the review in 
Ref.~\cite{PS185}. For the reaction $\pbarp \to \sbarl + c.c.$ the situation
is already less satisfactory and for 
$\pbarp \to \spbarsp$ and $\pbarp \to \smbarsm$ basically only a single
measurement exists. In case of $\pbarp \to \xbarx$ only upper bounds of the
reaction cross section are available. 
The quality of the description of the PS185 data by the J\"ulich model
is documented in 
Refs.~\cite{Haidenbauer:1991,Haidenbauer:1992,Haidenbauer:1993}. 

Since the reaction $\ebare\to \sbars$ is discussed in greater detail
in the present study 
we display in Fig.~\ref{fig:SSe} results of the employed $\ybary$ model 
\cite{Haidenbauer:1993} for $\pbarp \to \sbars$. Another reason is that 
at the time when the model was published only a (preliminary) data point
for the $\spbarsp$ channel was available. The results for 
$\pbarp \to \smbarsm$ were, therefore, genuine predictions. It is 
re-assuring to see that the model predictions are well in line with 
the final data, in particular, in view of the fact that $\pbarp \to \smbarsm$
requires double-charge exchange, i.e. involves at least a two-step process.
\begin{figure}[htb!]
\begin{center}
\includegraphics[height=85mm,angle=-90]{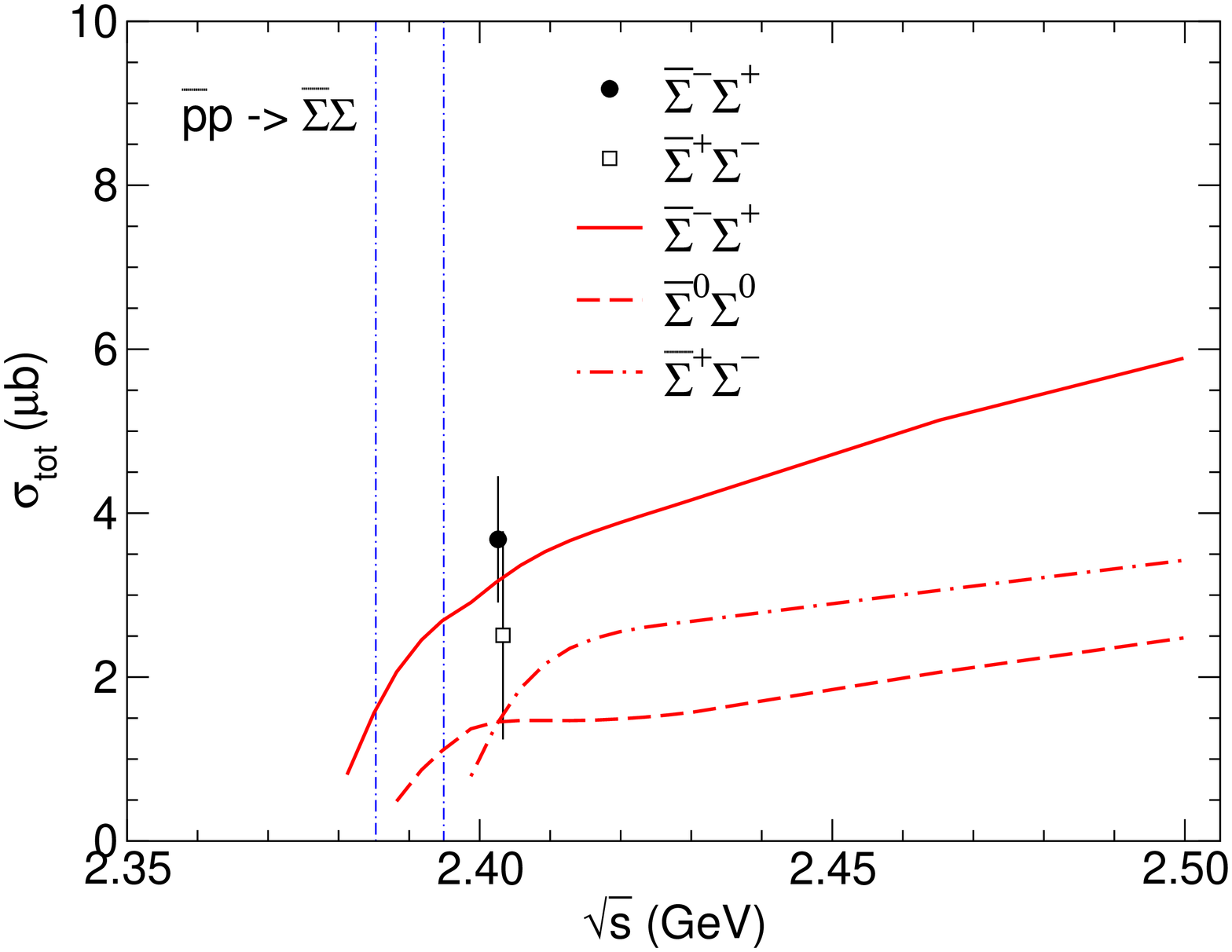}
\caption{Results for $\pbarp \to \sbars$ based on the $\ybary$ model presented 
in Ref.~\cite{Haidenbauer:1993}. 
Data are taken from Refs.~\cite{Barnes:1997,Johansson:1999,TordSS}. 
The horizontal dash-dotted lines indicate the $\sobarso$ and $\smbarsm$ thresholds.
}
\label{fig:SSe}
\end{center}
\end{figure}
 
Partial wave cross sections at the excess energy of $\epsilon = 24$~MeV are listed 
in Table~3 of Ref.~\cite{Haidenbauer:1993}.
One can see that the by-far largest contribution comes from the $^3P_2$ partial
wave, despite of the low energy. The contribution of the $^3S_1$ amounts to 
only 30~\% ($\spbarsp$) and 10~\% ($\smbarsm$) of the $\pbarp\to\sbars$ 
transition cross section. In the meson-exchange potential the cross section is 
driven by the strong (and coherent) tensor forces due to $K$ and $K^*$ 
exchange which leads to the characteristic dominance of $L \to L-2$ transitions 
\cite{Haidenbauer:1993}. This will be certainly different for $\ebare\to\sbars$
where the prime mechanism is $s$-channel one-photon exchange. 

Predictions of the model for the total $\sbars$ cross sections and 
for $\lbarl\to \sbars$ and $\lbars\to \sbars$ transition cross sections can 
be found in Figs.~23 and 24 of Ref.~\cite{Haidenbauer:1993}.
Note that the transition cross sections are shown in $\mu b$ and suggest 
that the coupling to $\sbars$ is relatively week. In view of that, we do 
not take into account those channels
explicitly in the evaluation of the $\sbars$ form factors via Eq.~(\ref{FSI}).
Rather we assume that their effect can be absorbed into the ``bare'' form factors, 
which anyway involve free parameters that need to be determined by a fit to 
the BESIII data.

The limited amount of data for $\pbarp \to \sbars$ did not require a refined 
adjustment of the $\sbars$ interaction. Specifically, the annihilation part
is described only by a central potential. Unlike the situation for $\lbarl$
and $\sbarl$ no LS or tensor component was used because the corresponding strength 
and range parameters could not be fixed form the single data point available
at that time \cite{Haidenbauer:1993}. 


\begin{figure}[t]
\begin{center}
\includegraphics[height=85mm,angle=-90]{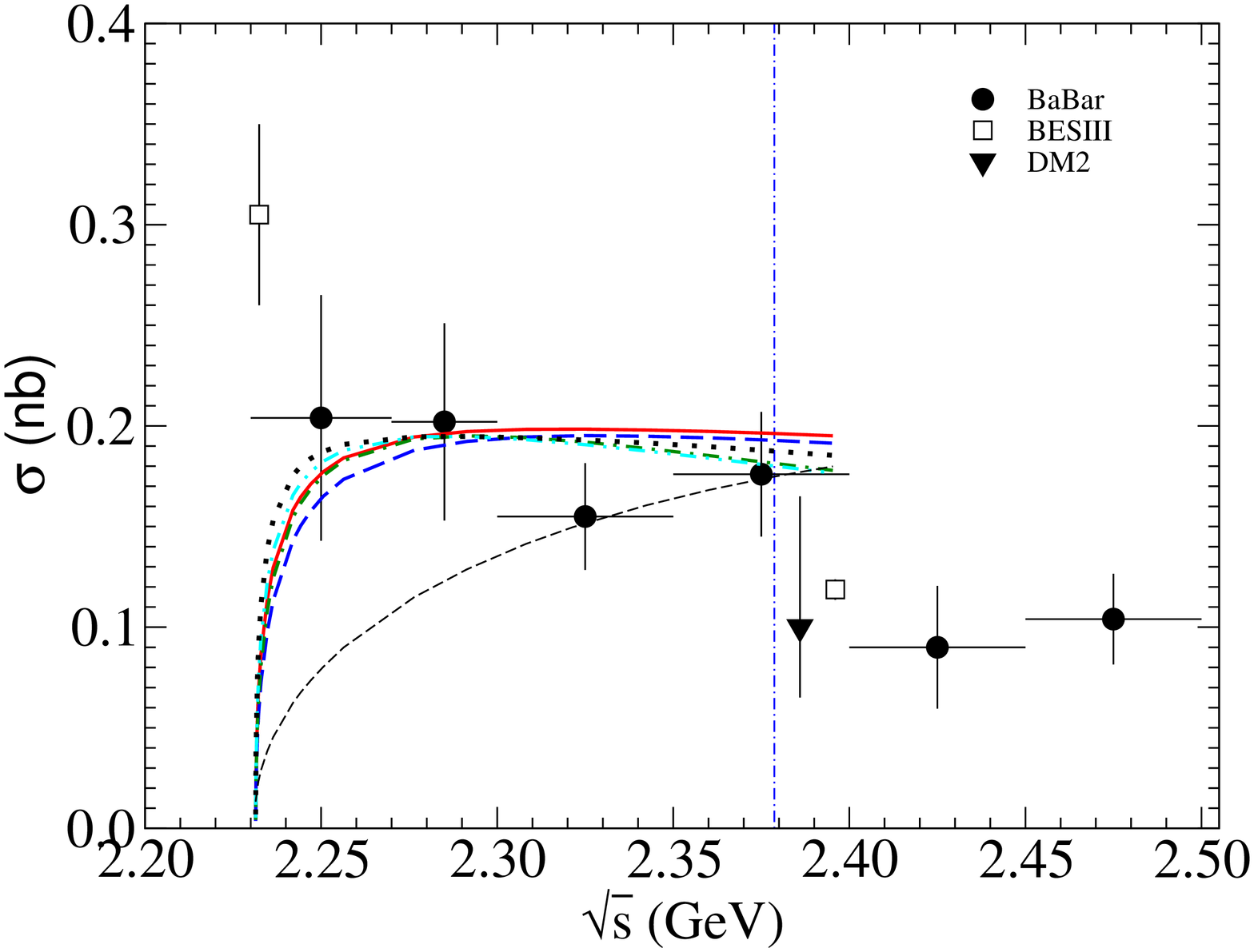}\includegraphics[height=85mm,angle=-90]{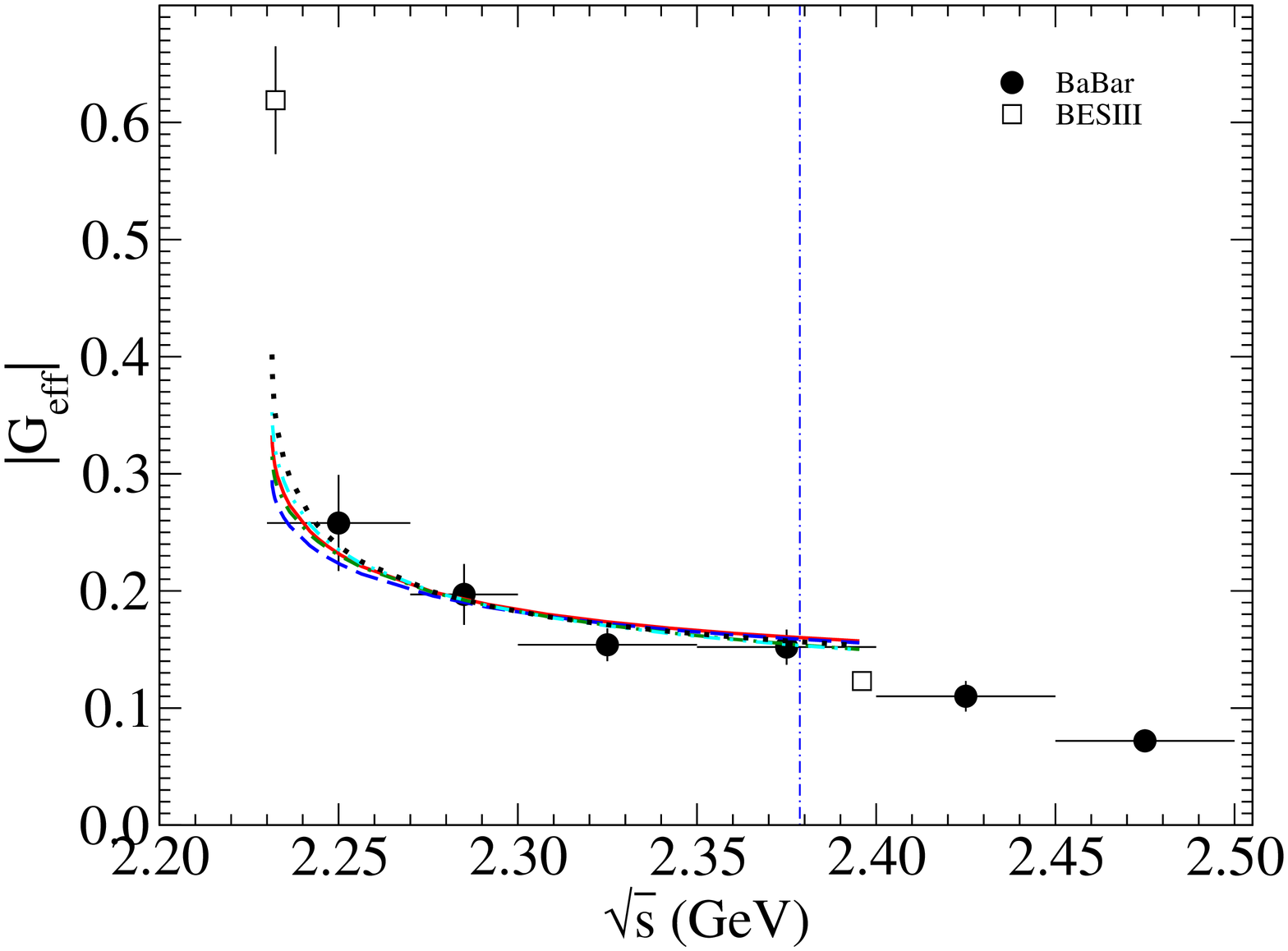}
\caption{$\ebare \rightarrow \lbarl$:
Cross section (left) and effective form factor $|G_{\rm eff}|$ (right)
as a function of the total cms energy.
The solid, dashed, and dash-dotted lines correspond to the $\lbarl$ models I, II, and III
from Ref.~\cite{Haidenbauer:1991}, the dash-double-dotted and dotted lines
to the models K and Q described in Ref.~\cite{Haidenbauer:1992}.
Data are from the Refs.~\cite{Bisello:1990} (DM2), \cite{Aubert:2007} (BaBar), and
\cite{Ablikim:2017,Ablikim:2019} (BESIII).
The phase space is shown by the thin (black) dashed line. 
The horizontal dash-dotted line indicates the isospin-averaged $\sbars$ threshold.
}
\label{fig:LL}
\end{center}
\end{figure}

\begin{figure}[htb!]
\begin{center}
\includegraphics[height=85mm,angle=-90]{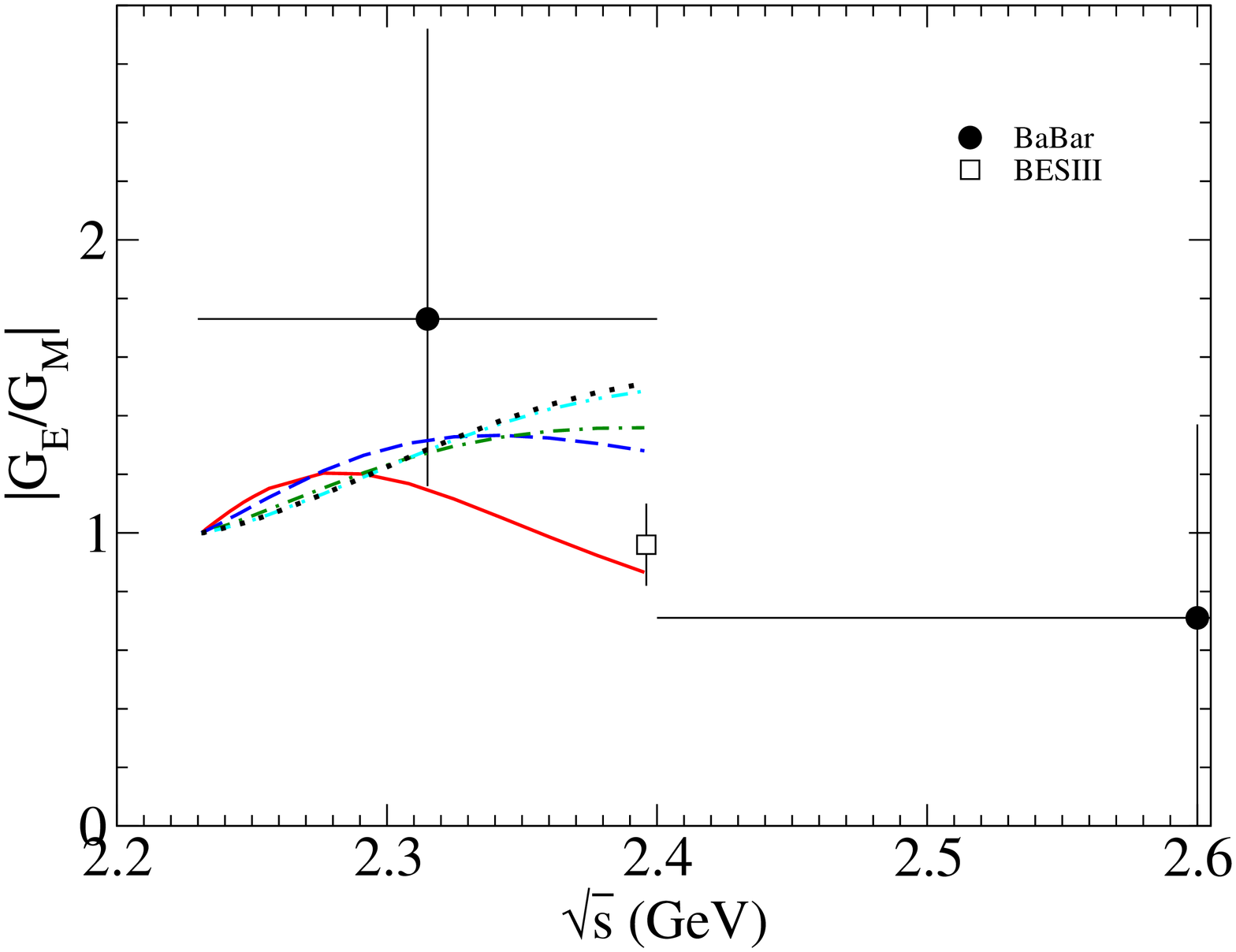}\includegraphics[height=85mm,angle=-90]{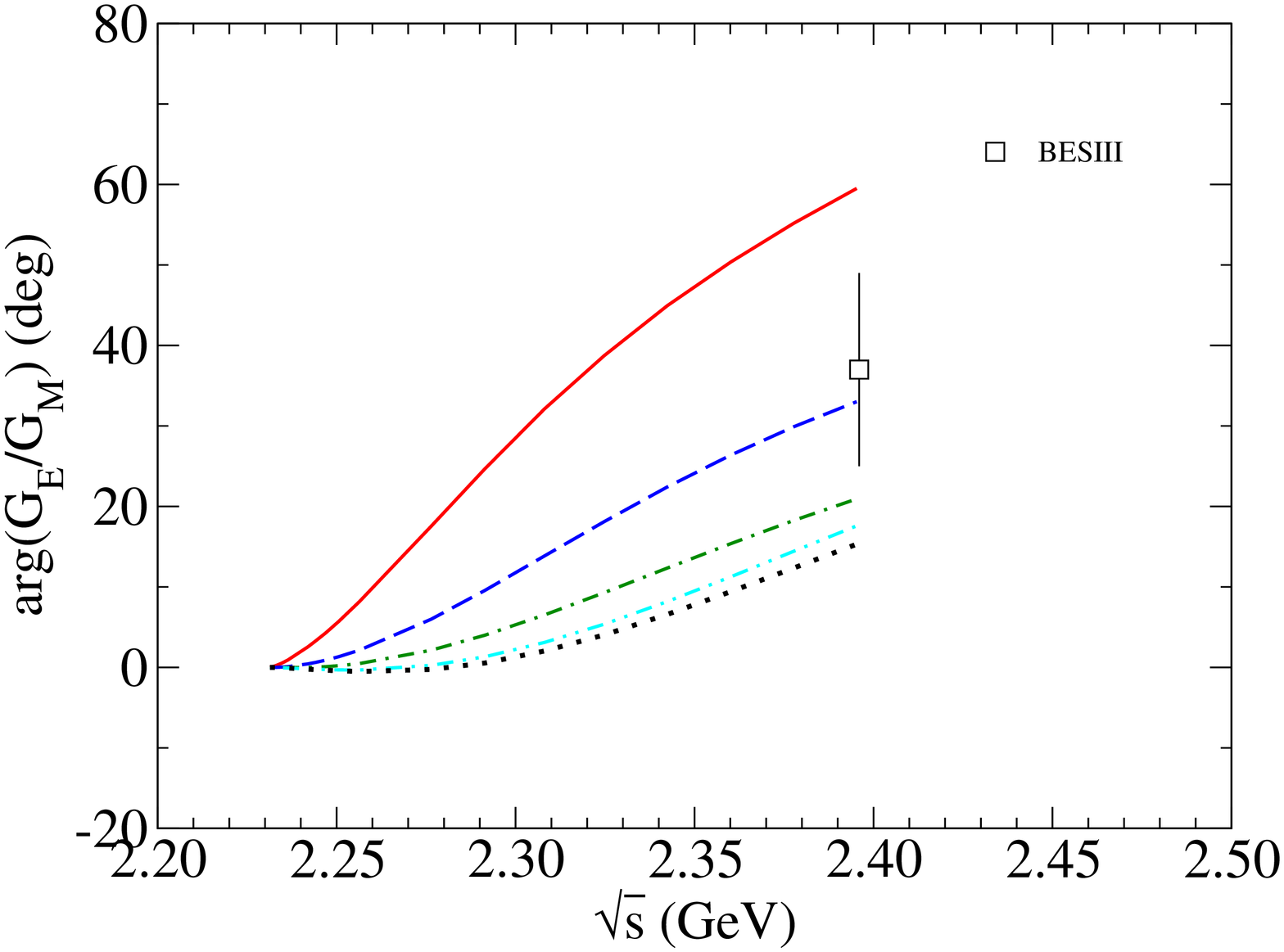}
\caption{$\ebare \to \lbarl$: 
The ratio $|G_E/G_M|$ (left) and phase $\phi=\rm{arg}(G_E/G_M)$ (right) as a function of the
total cms energy. 
Same description of curves as in Fig.~\ref{fig:LL}.
Data are from Refs.~\cite{Aubert:2007} (BaBar) and \cite{Ablikim:2019} (BESIII).
}
\label{fig:LLGEM}
\end{center}
\end{figure}

\section{Results for {\boldmath$\ybary$}}
\label{sec:results}

\subsection{{\boldmath$\ebare \to \lbarl$}}

Our results for $\ebare \to \lbarl$ have been discussed in detail in Ref.~\cite{Haidenbauer:2016}.
We reproduce some of them here because at the time when our work was published only 
preliminary results from the BESIII Collaboration were available. 
In the meantime final results have been presented \cite{Ablikim:2017}. As can be seen in 
Fig.~\ref{fig:LL} the initially reported rather large cross section extremely 
close to the threshold has been confirmed. Its value is not in line with the 
trend suggested by the BaBar data and, also, it cannot be reproduced by our $\lbarl$ FSI.
A plausible and convincing explanation/interpretation is still missing  
\cite{Cao:2018,Yang:2018,Yang:2019,Xiao:2019}.
For a more thorough discussion of this issue see Ref.~\cite{Haidenbauer:2016}.
Interestingly, the BESIII Collaboration has accomplished a complete measurement of the $\La$ 
EMFFs at $2.396$~MeV, which produced experimental values for the ratio $|G_E/G_M|$ and for the 
phase between $G_E$ and $G_M$, i.e. $\phi=\rm{arg}(G_E/G_M)$ \cite{Ablikim:2019}. 
Those data are confronted with our predictions in Fig.~\ref{fig:LLGEM}.
Evidently, among the various models considered in our study \cite{Haidenbauer:2016}, the 
results based on model~I from Ref.~\cite{Haidenbauer:1991} are closest to the measurement.
Note that $|G_E/G_M|$ is shown for an expanded energy range so that one can see better the
trend for this observable as suggested by the BaBar data point at $2.6$~GeV.

\begin{figure}[htb!]
\begin{center}
\includegraphics[height=85mm,angle=-90]{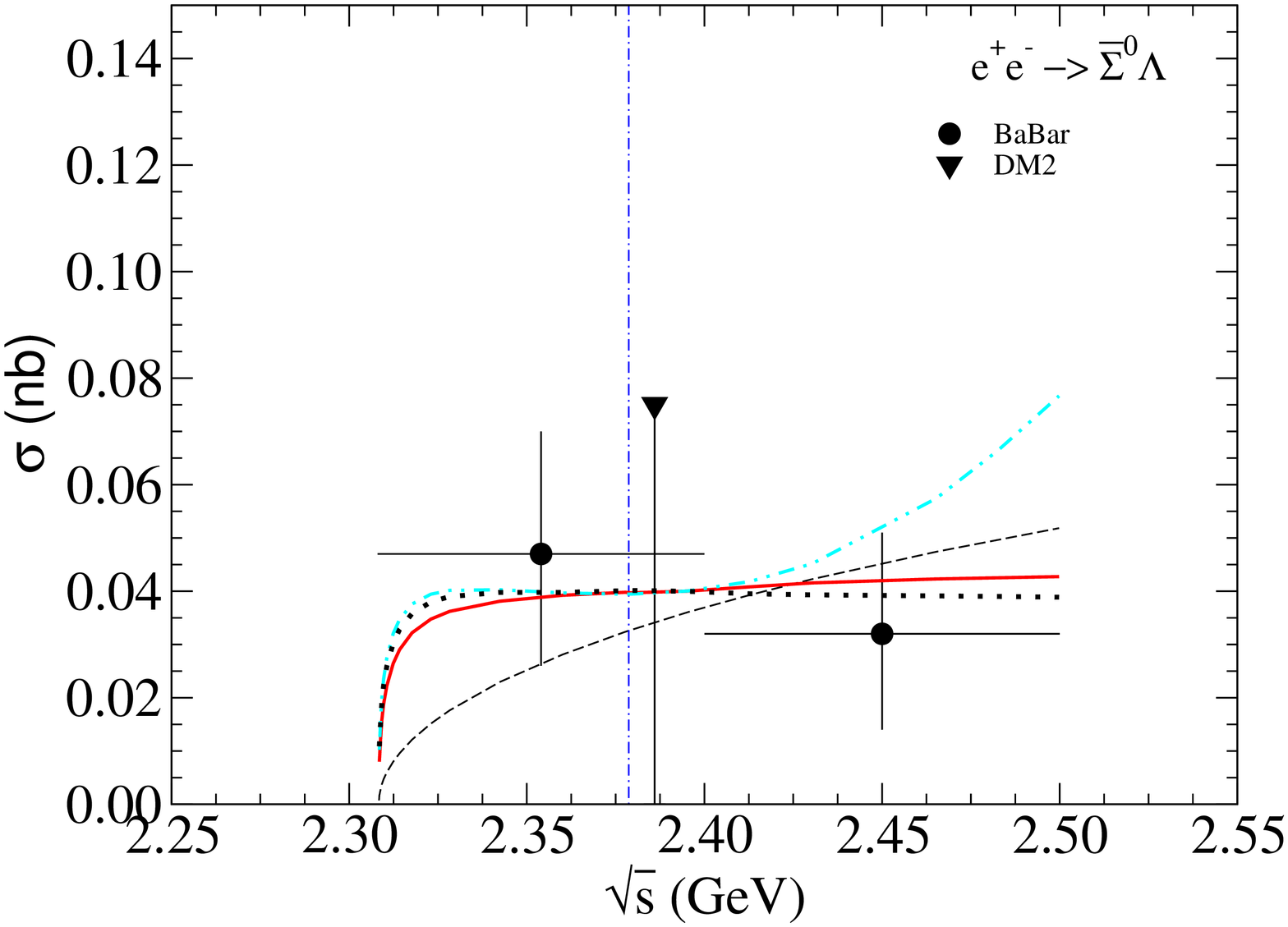}\includegraphics[height=85mm,angle=-90]{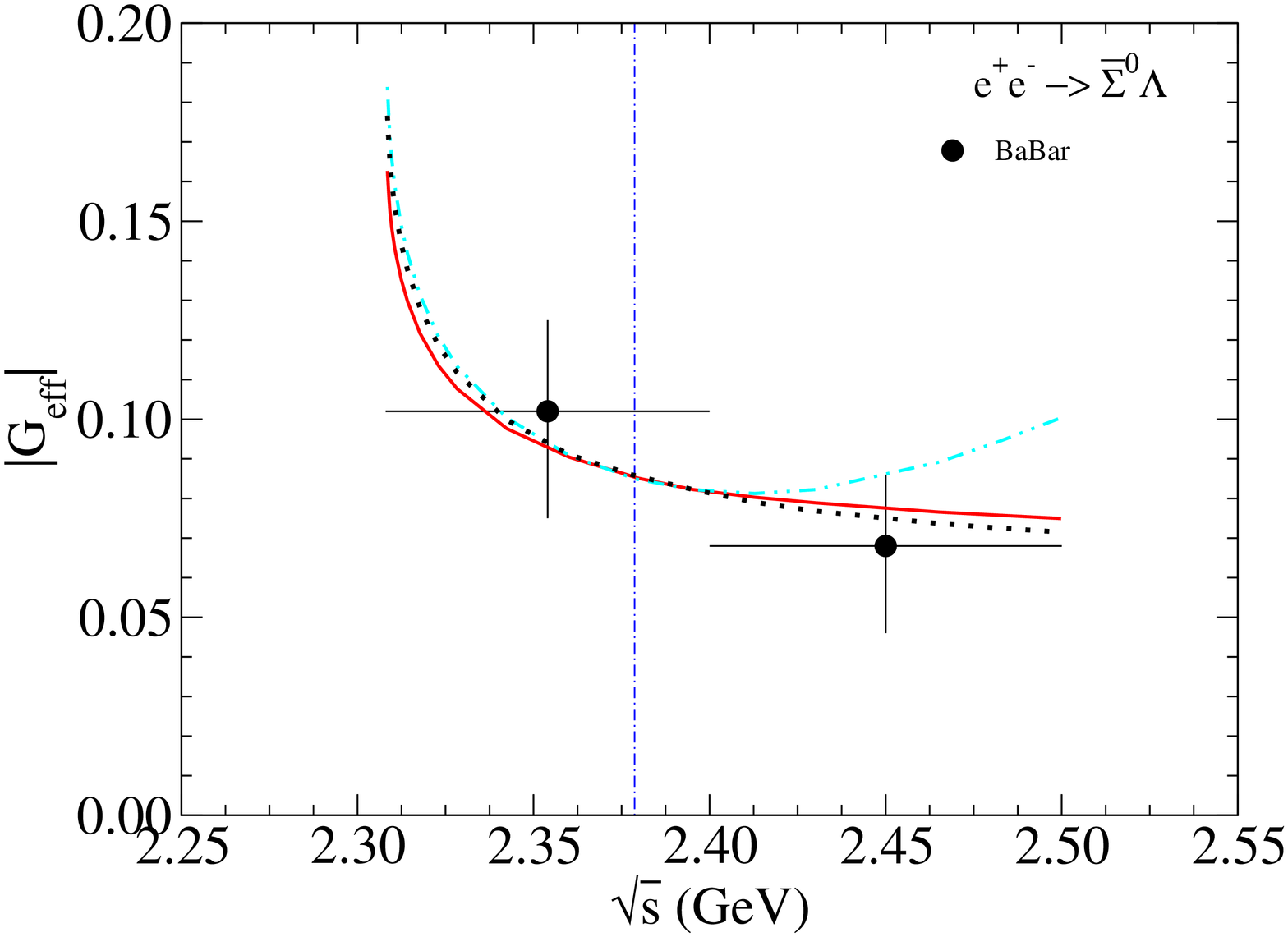}
\caption{$\ebare \to \sbarl$: Cross section (left) and effective form factor $|G_{\rm eff}|$ (right).
The solid, dash-double-dotted and dotted lines correspond to the models I, K and Q described in 
Ref.~\cite{Haidenbauer:1993}.
Data points are taken from the DM2 \cite{Bisello:1990} and
BaBar~\cite{Aubert:2007} Collaborations.
The phase space is shown by the thin (black) dashed line. 
The horizontal dash-dotted line indicates the $\sbars$ threshold.
}
\label{fig:LS}
\end{center}
\end{figure}

\begin{figure}[htb!]
\begin{center}
\includegraphics[height=85mm,angle=-90]{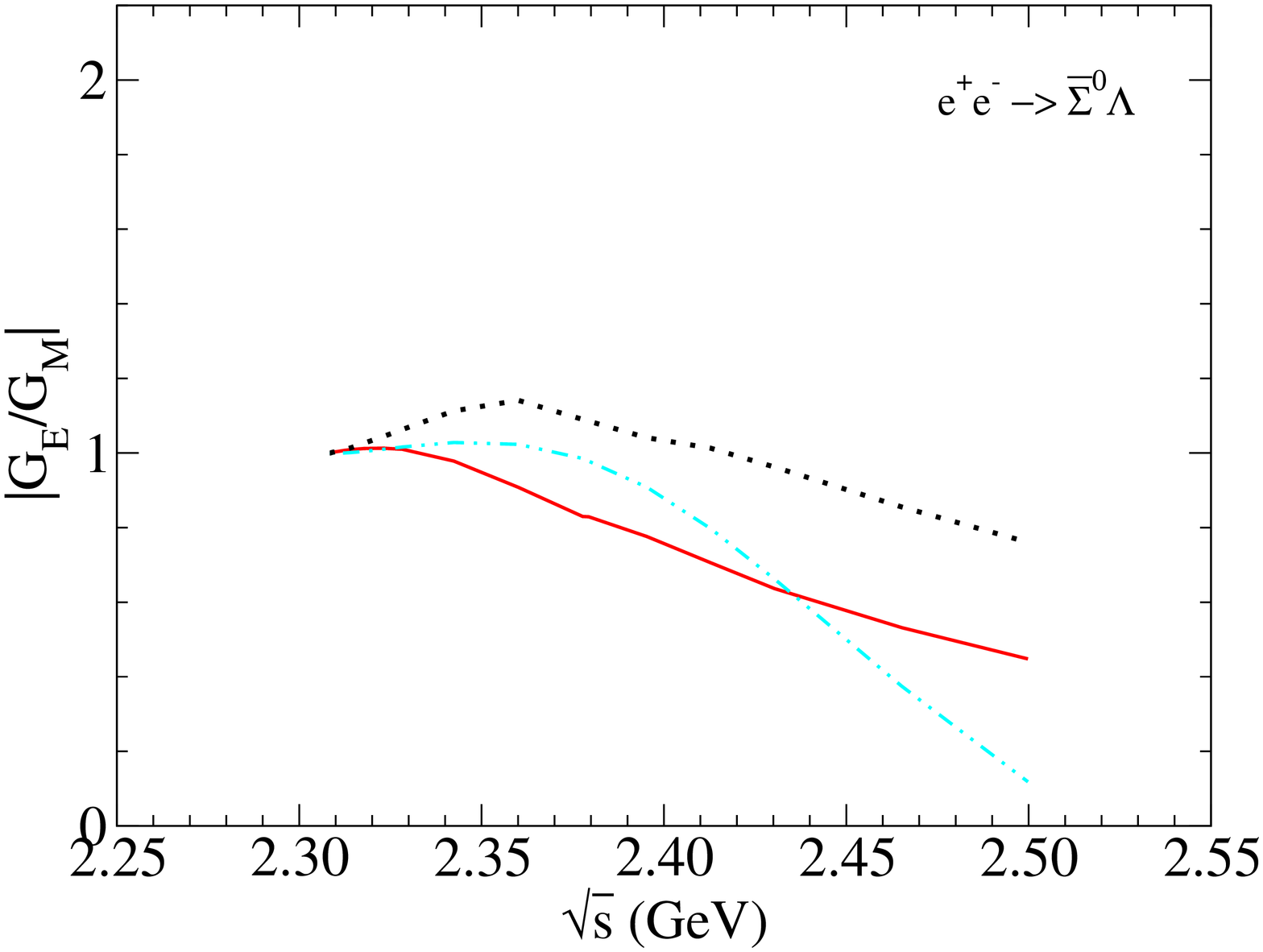}\includegraphics[height=85mm,angle=-90]{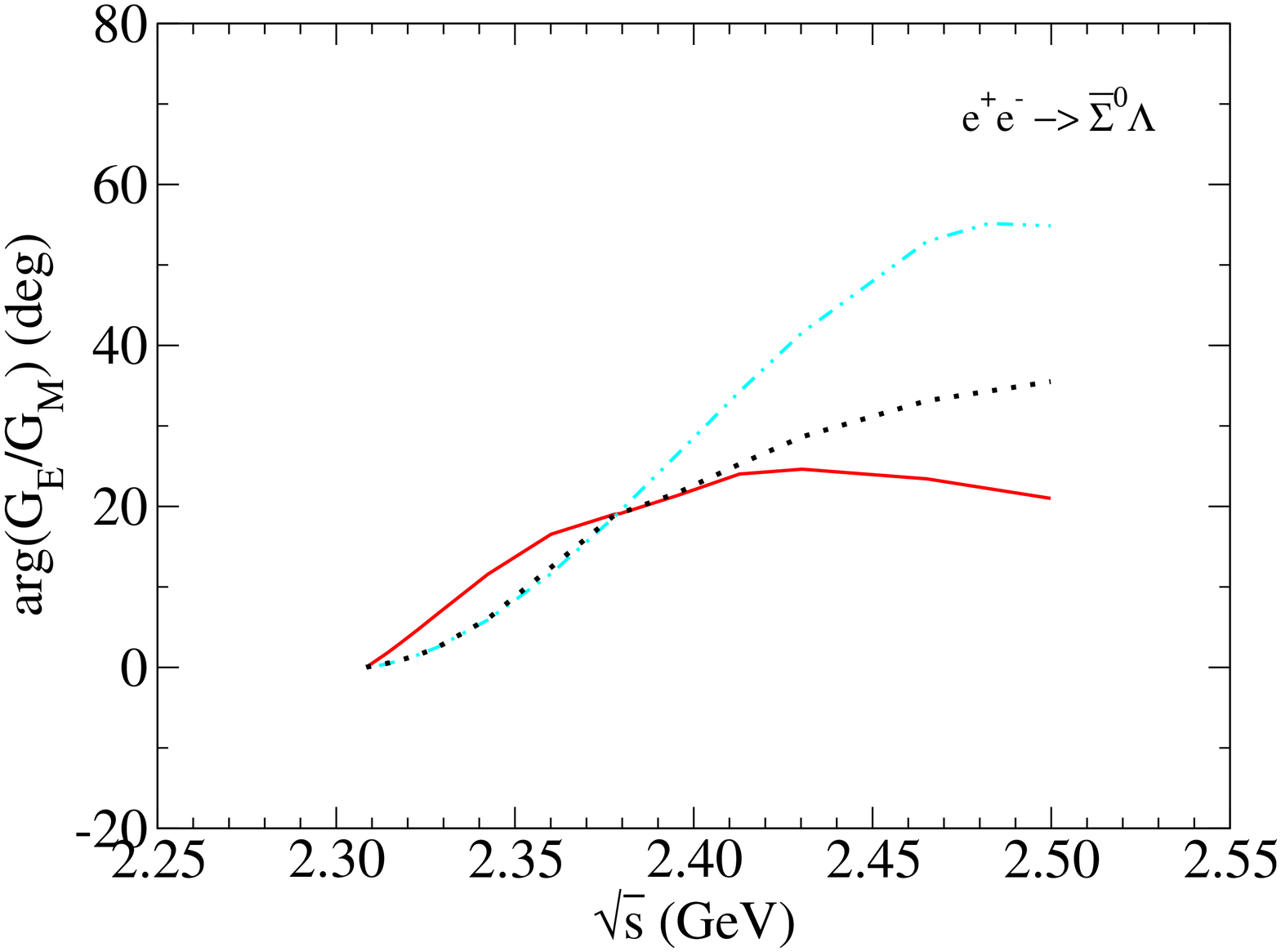}
\caption{$\ebare \to \sbarl$: Ratio $|G_E/G_M|$ (left) and phase $\phi=\rm{arg}(G_E/G_M)$ (right). 
Same description of curves as in Fig. ~\ref{fig:LS}. 
}
\label{fig:LSGEM}
\end{center}
\end{figure}

\subsection{{\boldmath$\ebare \to \sbarl$}}

Results for $\ebare \to \sbarl$ are presented in Figs.~\ref{fig:LS} and \ref{fig:LSGEM}. 
For that reaction only cross sections are avaible \cite{Aubert:2007}. The data point 
from DM2 \cite{Bisello:1990} is an upper limit. 
We fixed the normalization (i.e. the bare form factor) so that the cross sections are 
$0.04$~nb at $2.4$~GeV for all three $\sbarl$ models presented in \cite{Haidenbauer:1993}.
Comparing Figs.~\ref{fig:LS} and \ref{fig:LL} one can see that 
the $\sbarl$ cross section is about a factor $5$ smaller than that for $\lbarl$. 
However, like the latter one it remains practically constant over a fairly large energy 
region above the threshold and does not follow the phase space behavior. 
This conclusion can be savely drawn despite of the modest energy resolution. 
The results based on two of the $\ybary$ models considered in 
Ref.~\cite{Haidenbauer:1993} agree nicely with that behavior over the whole considered
energy region. One of the models suggest a rise in the cross section from around 
$2.4$~GeV onwards which is clearly ruled out by the experiment. 
As can be seen in Fig.~\ref{fig:LSGEM}, 
also for the ratio $|G_E/G_M|$ and the phase $\phi=\rm{arg}(G_E/G_M)$ 
a qualitative similar behavior to that for the $\lbarl$ channel is predicted.

\begin{figure}[t]
\begin{center}
\includegraphics[height=85mm,angle=-90]{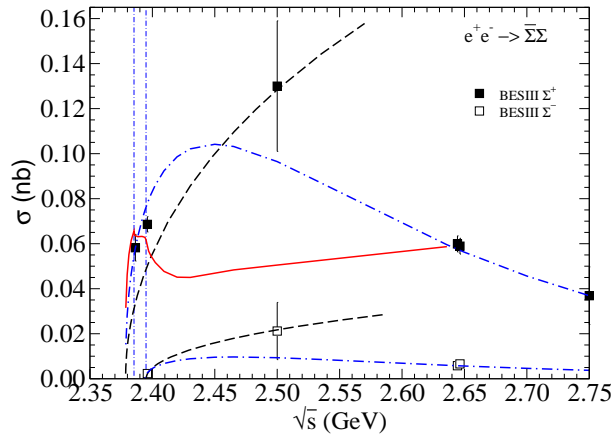}
\caption{$\ebare \to \sbars$: Illustrative curves/results for the 
$\spbarsp$ and $\smbarsm$ cross sections.
The dashed lines represent the phase space behavior. 
The solid line is the result of an exploratory calculation that uses only the 
$\spbarsp\to\spbarsp$ amplitude for the FSI.  
The dash-dotted line represents the pQCD motivated parameterization Eq.~(3) 
in \cite{Ablikim:2020}. Data are taken from Ref.~\cite{Ablikim:2020}.
The horizontal lines indicate the $\sobarso$ and $\smbarsm$ thresholds.
}
\label{fig:SSi}
\end{center}
\end{figure}

\subsection{{\boldmath$\ebare \to \sbars$}}

Let us begin the discussion on the $\sbars$ channel with some general considerations.
Alternatively to \cite{Ablikim:2020}, in Fig.~\ref{fig:SSi} we present the BESIII results
up  to $2.75$~GeV on a linear scale. Data for $\spbarsp$ (filled squares)
and $\smbarsm$ (opaque squares) are included together with their statistical uncertainties.
The systematic uncertainties quoted in the paper are usually of the same
magnitude \cite{Ablikim:2020}.  
The pure phase space behavior corresponds to the dashed lines. 
The results based on the pQCD motivated 
fit via Eq.~(3) in Ref.~\cite{Ablikim:2020} are indicated by the dash-dotted lines. 
The result of an exploratory calculation based on the $\ybary$ model from 
Ref.~\cite{Haidenbauer:1993}, where only the $\spbarsp$ amplitude is taken 
into account in Eq.~(\ref{FSI}), is shown by the solid line. 
The horizontal dashed lines indicate the $\sobarso$ and $\smbarsm$ thresholds. 

Evidently, the first two data points for $\spbarsp$ are close to the
thresholds of the $\sobarso$ and $\smbarsm$ channels, respectively. 
Furthermore, the first three data points are roughly in line with the
phase-space behavior, if one takes into account the large uncertainty of the
measurement at $\sqrt{s}=2.5$~GeV. At first sight this suggest that, unlike what
was observed for $\lbarl$ and $\sbarl$, FSI effects could be very small in case of
$\sbars$. However, the next two data points, around $2.65$~GeV, suggest a quite
different energy dependence. These cross sections are of comparable magnitude to 
the ones around $2.39$~MeV and would be in line with a basically flat 
behavior similar to what has been seen for $\lbarl$ and $\sbarl$. 
We include here the results based on the pQCD inspired function considered
in the BESIII paper, cf. Eq.~(3) in Ref.~\cite{Ablikim:2020}. It allows one to reproduce the 
energy dependence of the cross section quite well over the whole energy range,
except for the data point at $2.5$~GeV. 
The situation for the $\smbarsm$ channel is similar. Also here the 
cross section at $2.5$~GeV is not described. 

We take the above observations as a strong indication against using the 
cross sections at $2.5$~GeV for fixing the parameters of our bare form 
factors (\ref{FOFA0}). Instead we utilize those around $2.65$~GeV. 
We want to emphasize that this choice is certainly debatable.
The two points in question are already more than $200$~MeV away from the $\sbars$ 
thresholds and, therefore, our assumption that the bare form factors are 
constant is most likely not valid over such a large energy range. 
One should keep this caveat in mind 
when assessing the results that will be presented below. 

\begin{figure}[t]
\begin{center}
\includegraphics[height=85mm,angle=-90]{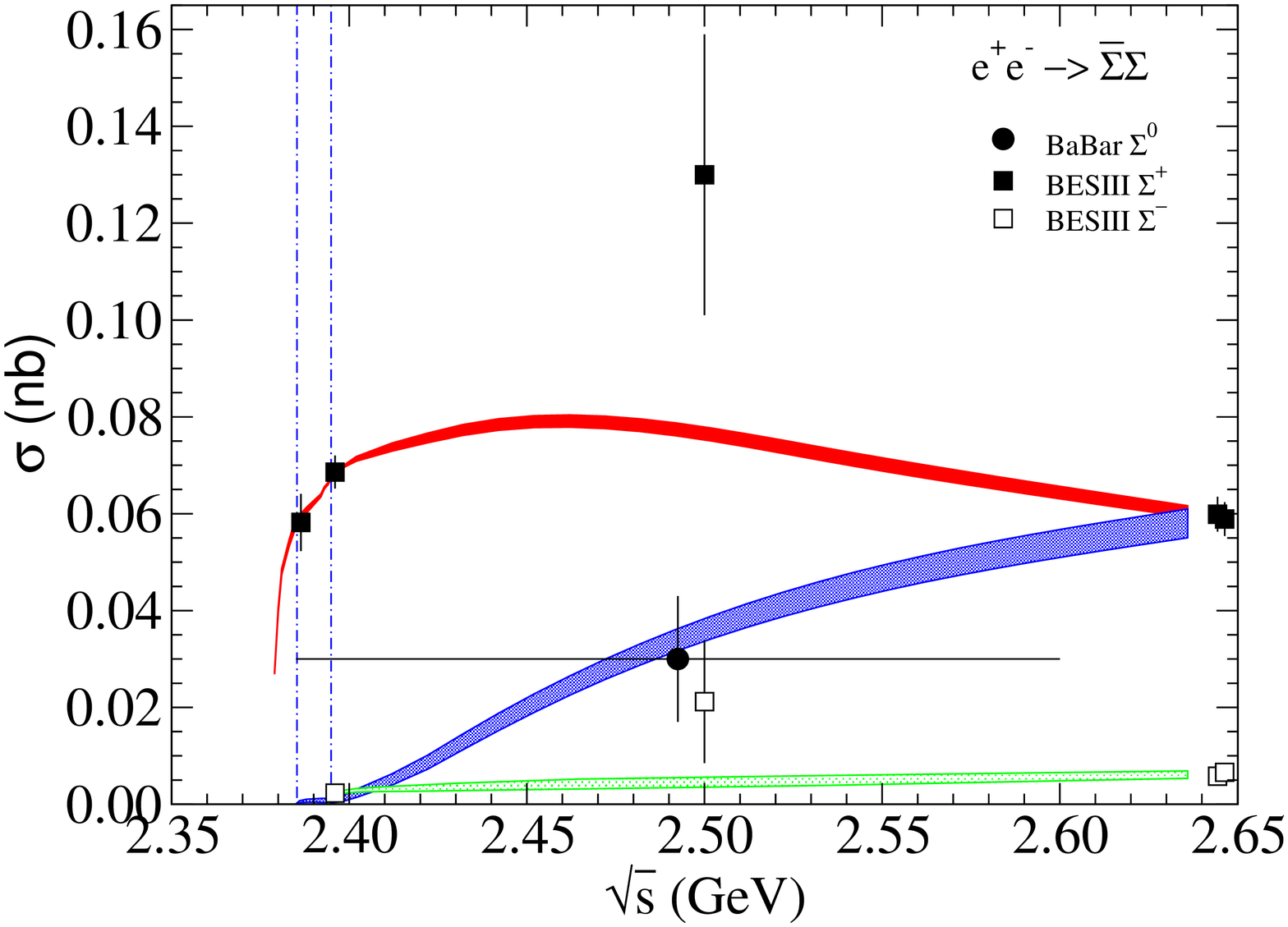}\includegraphics[height=85mm,angle=-90]{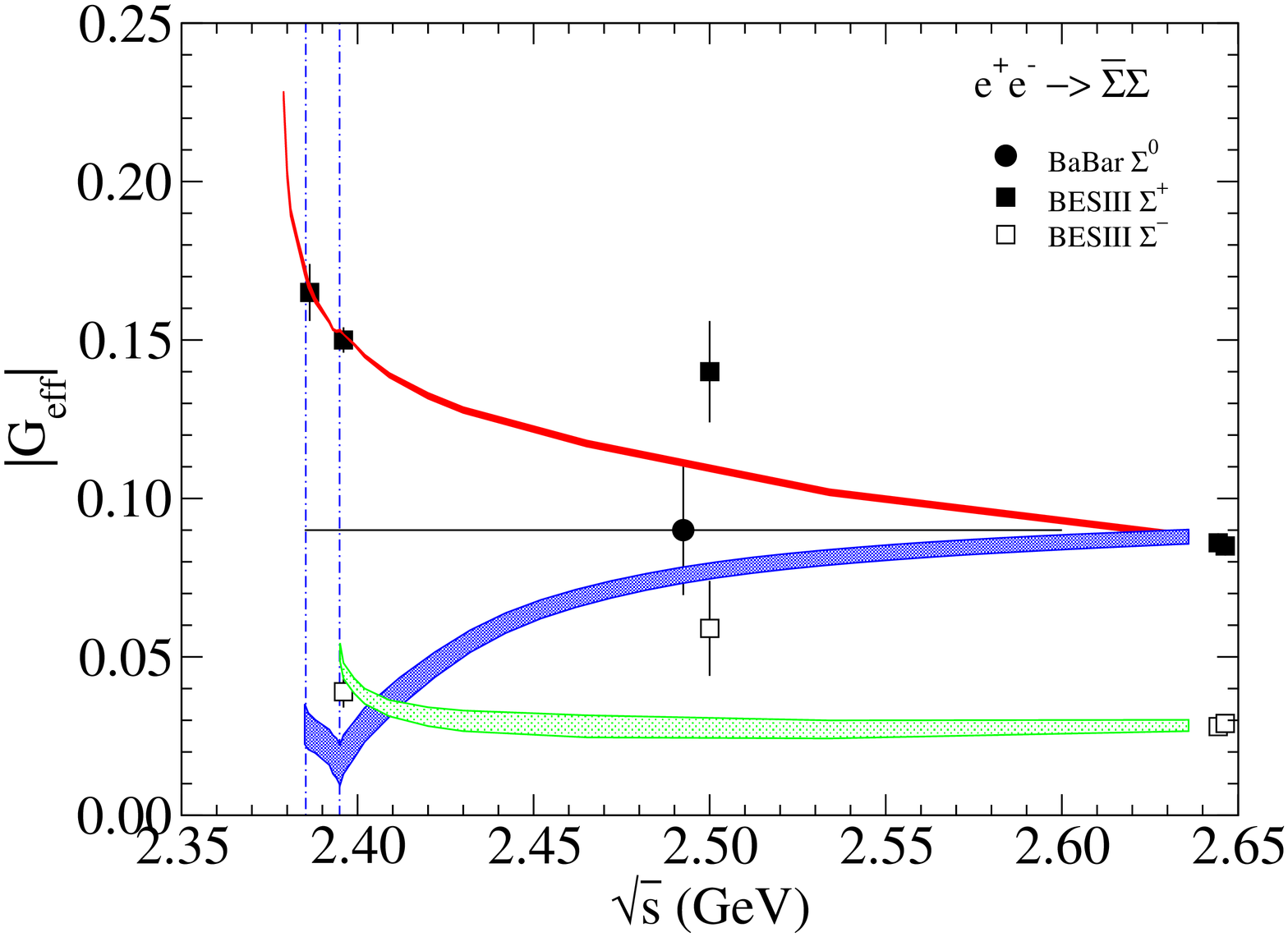}
\caption{$\ebare \to \sbars$: 
Cross section (left) and effective form factor $|G_{\rm eff}|$ (right).
Results are shown based on the $\sbars$ interaction from Ref.~\cite{Haidenbauer:1993}, 
where the dark (red) bands correspond to $\spbarsp$, 
medium (blue) bands to $\sobarso$, and light (green) bands to $\smbarsm$. 
For the meaning of the bands see text. 
Data for the $\spbarsp$ and $\smbarsm$ channels are from BESIII \cite{Ablikim:2020},  
those for $\sobarso$ from BaBar \cite{Aubert:2007}.
The horizontal dash-dotted lines indicate the $\sobarso$ and $\smbarsm$ thresholds.
}
\label{fig:SS}
\end{center}
\end{figure}

In the actual fitting process we also ignored the measured ratio $|G_E/G_M|$ 
at $2.396$~GeV. Here the BESIII Collaboration found a value of $1.83\pm 0.26$
\cite{Ablikim:2020} which is extremely large given that this energy is just 
about $17$~MeV away from the $\spbarsp$ threshold. Since $G_E = G_M$ at
the threshold, as mentioned above, one would need an excessively large 
$D$-wave component in order to achieve such a value, which cannot be 
generated within the conventional dynamics of the employed $\ybary$ model. 
Exotic contributions like near-threshold $D$-wave resonances might provide 
such drastic effects, but we do not want to be too speculative at this stage. 

Before proceeding to the actual fit it is worthwhile to look at the exploratory
calculation with just the $\spbarsp$ included (solid line in Fig.~\ref{fig:SSi}). 
The resulting cross section is normalized to the data points around $2.645$~GeV. Obviously,
then also the predictions around $2.39$~GeV are fairly well in line with
the experiment. However, more interesting is the energy dependence exhibited
in that region. One can clearly see threshold effects at the openings of
the $\sobarso$ as well as of the $\smbarsm$ channel. This is a clear signal
that the coupling between the $\sbars$ channels plays an important role for
the $\ebare$ cross sections, at least within the employed $\ybary$ model, 
and we ought to include it in the actual calculation. 

\begin{figure}[t]
\begin{center}
\includegraphics[height=85mm,angle=-90]{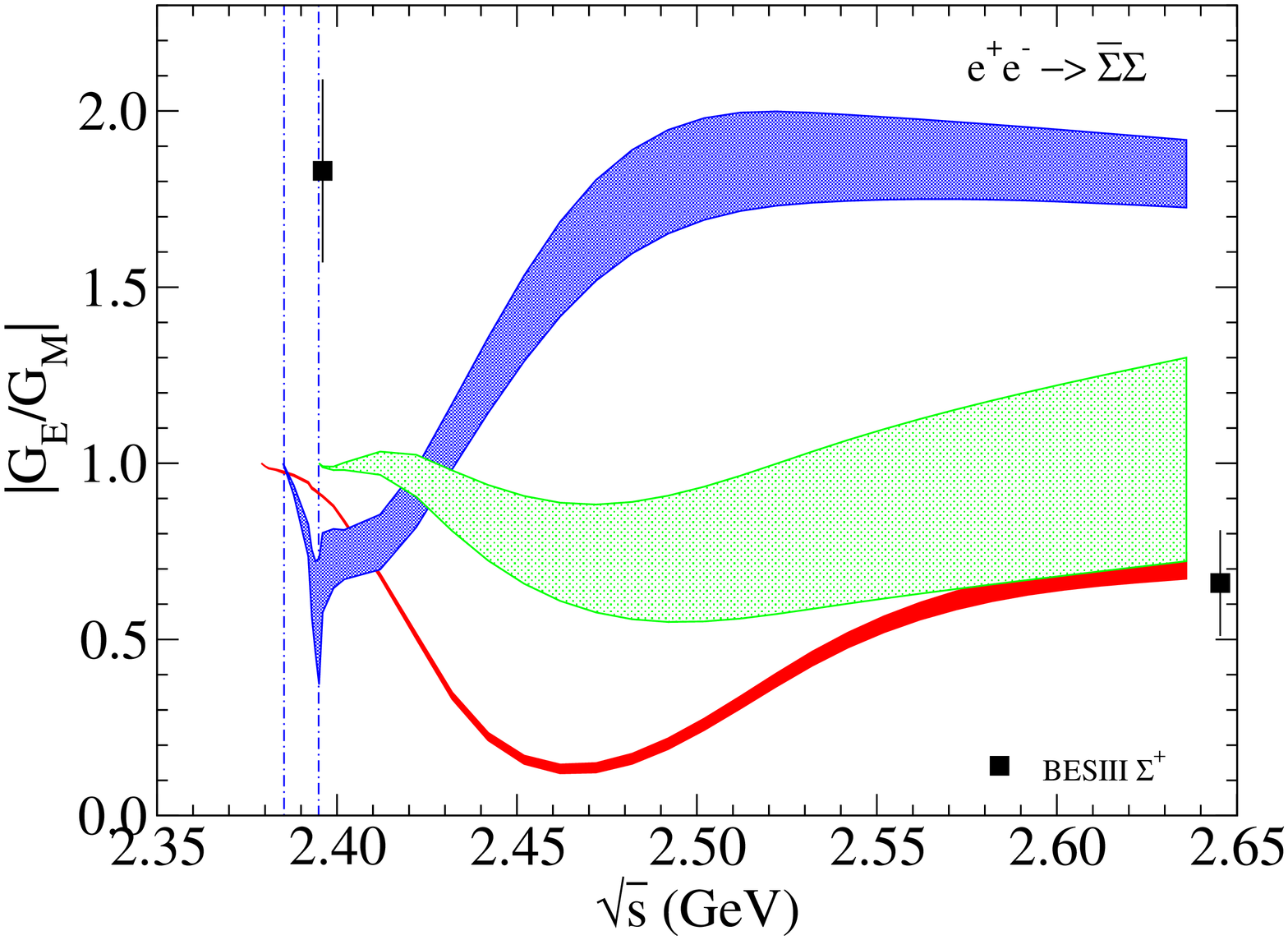}\includegraphics[height=85mm,angle=-90]{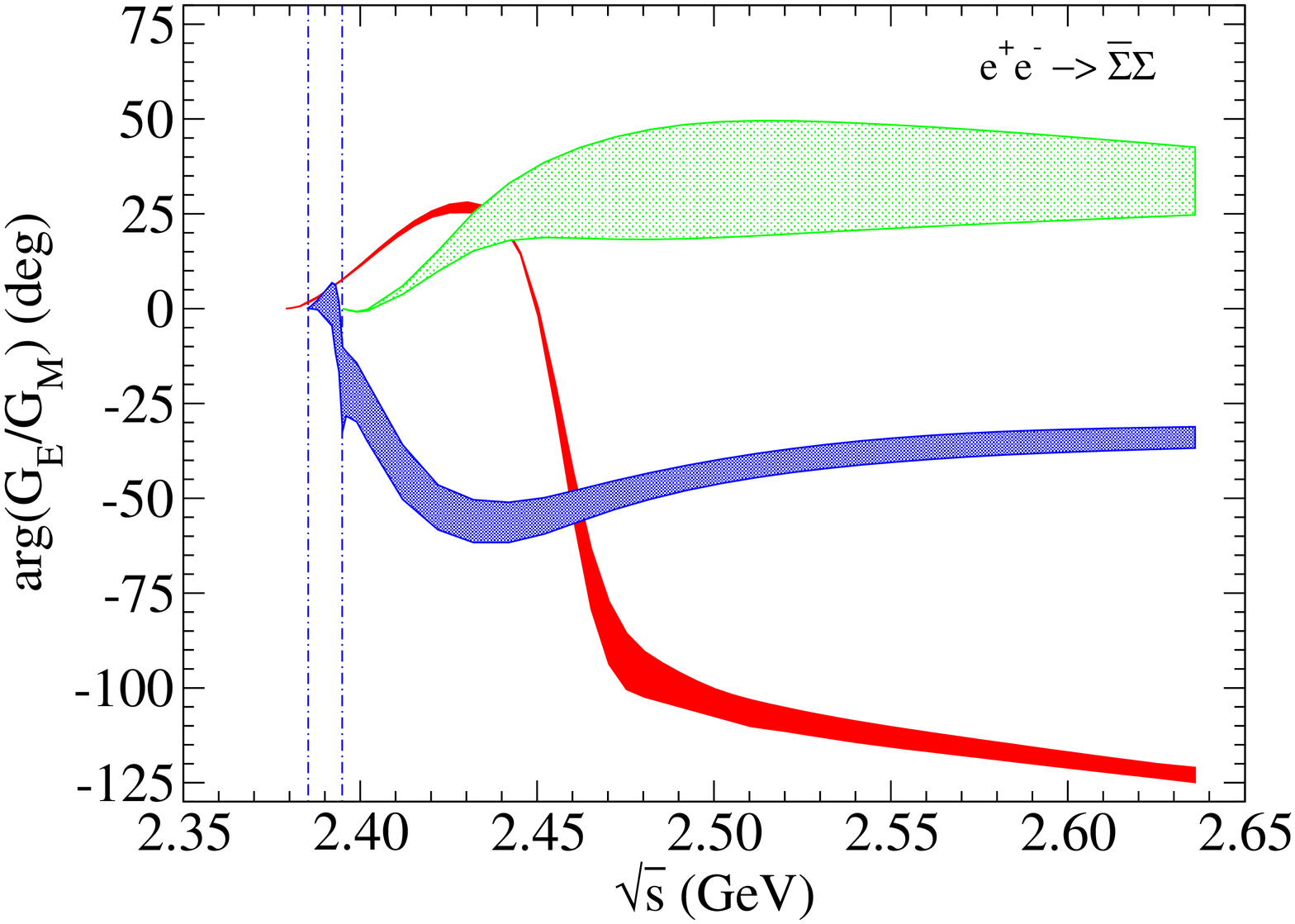}
\caption{$\ebare \to \sbars$: 
Ratio $|G_E/G_M|$ (left) and phase $\phi=\rm{arg}(G_E/G_M)$ (right).
Same description of curves as in Fig.~\ref{fig:SS}. 
Data for $\spbarsp$ are from the BESIII Collaboraton \cite{Ablikim:2020}.  
}
\label{fig:RaSS}
\end{center}
\end{figure}

Results of our fit are shown in Figs.~\ref{fig:SS} and \ref{fig:RaSS}. 
In this fit the cross sections at $2.3864$, $2.3960$, 
$2.6444$, and $2.6464$~GeV are included and in addition the
ratio $|G_E/G_M|$ at the latter two energies. 
The free parameters are the values of $G^0_M$ for 
$\spbarsp$, $\sobarso$, and $\smbarsm$ that enter the corresponding
vertex functions $f^0_0$ and $f^0_2$, see Eqs.~(\ref{FOFA0}). 
$G^{0\,(\Sigma^+)}_M$ is chosen to be real, since the overall phase
drops out in the calculation of the observables anyway, the other two 
are complex. Note that $G^{0\,(Y)}_M \equiv G^{0\,(Y)}_E$ so 
that there are no further parameters. 
The results are normalized to the $\spbarsp$ cross section at $2.3960$~GeV. 
However, we have checked that choosing a different energy for normalization 
leaves the results practically unchanged. 

As a consequence of the lack of data, only a single $\sbars$ interaction 
has been presented in Ref.~\cite{Haidenbauer:1993}. 
It turned out that a comparable description of the data points could 
be obtained for a range of parameters for the three bare form factors.  
Therefore, we present the achieved results as bands in Figs.~\ref{fig:SS} 
and \ref{fig:RaSS} which visualize this uncertainty. 
A typical example for the relative weights of the bare form factors as 
they come out of the fitting procedure is 
$\spbarsp$/$\sobarso$/$\smbarsm =
(1.00,0.00) / (0.55,0.30) / (0.00,0.60)$. 
Obviously the reproduction of the measured cross section ratio,
$\sigma_{\spbarsp}/\sigma_{\smbarsm}\approx 10$, requires that 
$G^{0\,(\Sigma^+)}_M$ and $G^{0\,(\Sigma^-)}_M$
are basically orthogonal. 

As one can see from Fig.~\ref{fig:SS}, the data points included in the fit are 
quite well described.
The $\spbarsp$ and $\smbarsm$ cross sections exhibit a weak energy dependence that 
is more moderate than the one of the ``pQCD'' parameterization \cite{Ablikim:2020}.  
This is not least due to the fact that the two near-threshold cross sections
at $2.3864$ and $2.3960$~GeV are quite well described, which is not the case for that
parameterization, see the dash-dotted line in Fig.~\ref{fig:SSi}. 
Interestingly, also the data point for $\sobarso$ from the BaBar Collaboration
\cite{Aubert:2007} is described, though it was not included in the fit. 
Admittedly, the energy resolution is much more modest than that of the BESIII data. 
Overall, however, the predicted energy dependence for $\sobarso$ differs from
the ones for the other two channel. 
The difference in the predicted energy dependence for $\sobarso$ can be seen 
likewise in the results for the effective form factors $|G_{\rm eff}|$ shown 
in the lower panel
of Fig.~\ref{fig:SS}. That figure reveals also another special feature of 
our $\sobarso$ result, namely a pronounced structure at the $\smbarsm$ threshold. 
In fact, there is also a structure in the $\spbarsp$ results at the 
$\sobarso$ and $\smbarsm$ thresholds. But those are rather moderate and, 
therefore, can be barely seen in the figure. 

In Fig.~\ref{fig:RaSS} predictions for the ratio $|G_E/G_M|$ and the
phase $\phi = {\rm arg} (G_E/G_M)$ are presented. Obviously these observables
reflect much more strongly the complicated dynamics in the coupled
$\sbars$ channels than the cross sections discussed above. It is also
worthwhile to recall that the same quantities show a fairly smooth
energy dependence in the $\lbarl$ and $\sbarl$ cases. 
Anyway, one can see that the ratio $|G_E/G_M|$ for $\spbarsp$ drops considerably
within the first $50$ MeV from the threshold and then increases again to reach
the (fitted) BESIII value of around $0.65$ at $2.64$~GeV. 
Even more spectacular is the behavior of the ratio in case of $\sobarso$ which
shows an extremely strong variation with energy within a very small energy region. 
In particular, there is a pronounced cusp effect at the opening of the $\smbarsm$.  
Clearly in view of that observation one can speculate that the large ratio 
measured by the BESIII Collaboration for $\spbarsp$ near the $\smbarsm$
threshold might be also a signal for such a cusp effect. The $\ybary$ 
interaction by the J\"ulich group does not give rise to a more pronounced effect 
in the $\spbarsp$ amplitude. However, as already said above, the rather limited
information on the $\pbarp\to\sbars$ reaction certainly did not allow to 
constrain the interaction in the $\sbars$ channel in that model in a reliable way. 
It goes without saying that more extended measurements in the region of
the $\sbars$ thresholds would definitely provide further insight into 
the $\ybary$ dynamics. In particular, it would be useful to explore the 
energy dependence of observables like the form factor ratio in detail. 

We have explored also alternative fit strategies. First we omitted the 
ratio $|G_E/G_M|$ for $\spbarsp$ at $2.64$~GeV. Then there is a stronger 
variation in the predictions for $|G_E/G_M|$, e.g. it is in the range of 
$0.7$ and $1.2$ at that energy, while the $\chi^2$ values for the cross sections 
remain basically the same. In addition, also the ratios for $\sobarso$ and $\smbarsm$ 
show larger variations.
Including the cross sections at $2.5$~GeV into the fit does not change the 
overall results much because of their comparatively large experimental uncertainties.   
Specifically, the energy dependence of the cross sections remain practically unchanged  
and only the overall $\chi^2$ increases noticeably. 

\begin{figure}[t]
\begin{center}
\includegraphics[height=85mm,angle=-90]{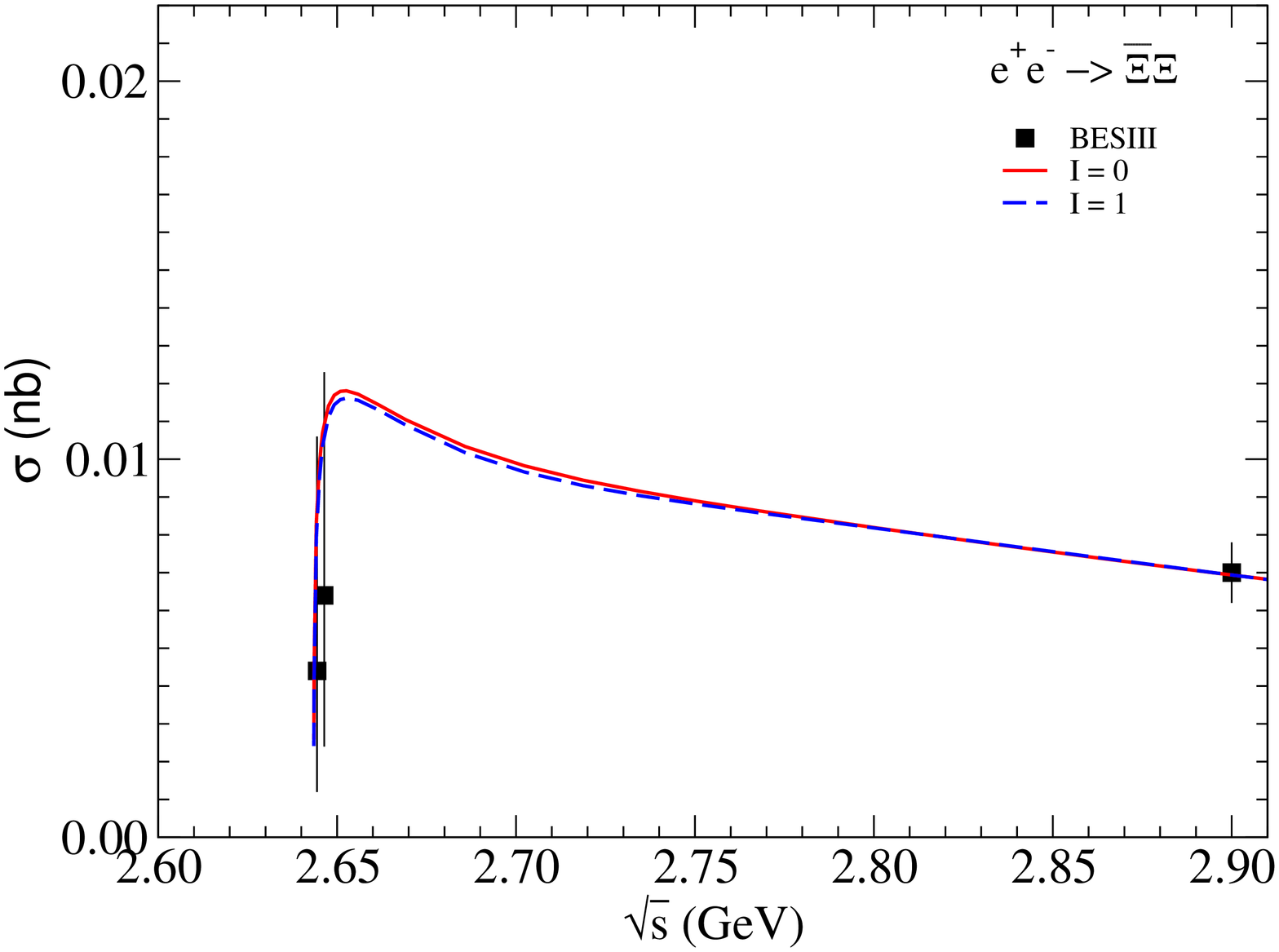}
\includegraphics[height=85mm,angle=-90]{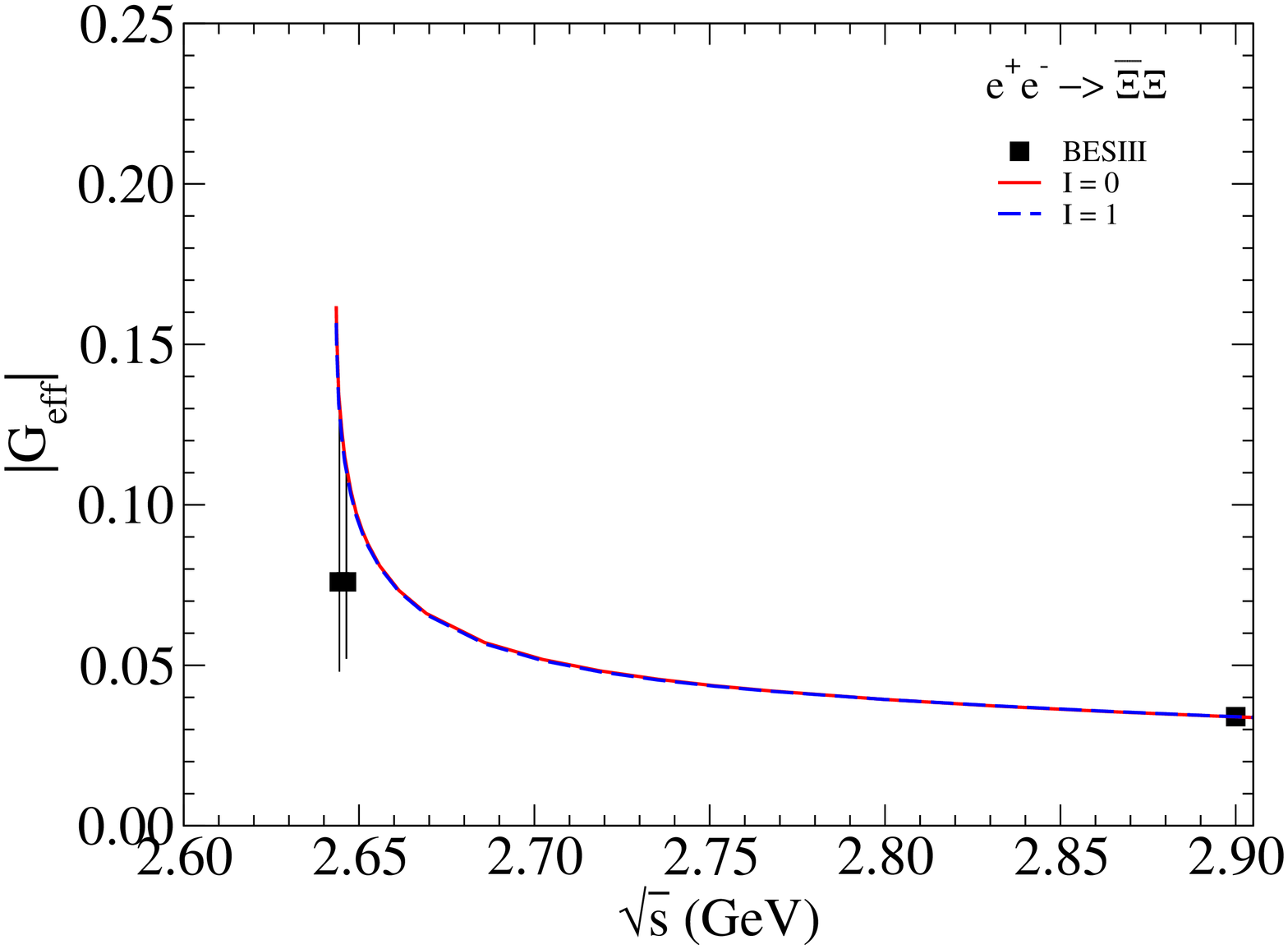}
\caption{$\ebare \to \xbarx$:
Cross section (left) and effective form factor $|G_{\rm eff}|$ (right).
Results for isospin $I=0$ (solid line) and $I=1$ (dashed line) 
are shown based on the $\xbarx$ interaction from Ref.~\cite{Haidenbauer:1993X}.
Data for the $\xmbarxm$ channel are from the BESIII Collaboration \cite{Ablikim:2020X2}.
}
\label{fig:XX}
\end{center}
\end{figure}
 
\begin{figure}[t]
\begin{center}
\includegraphics[height=85mm,angle=-90]{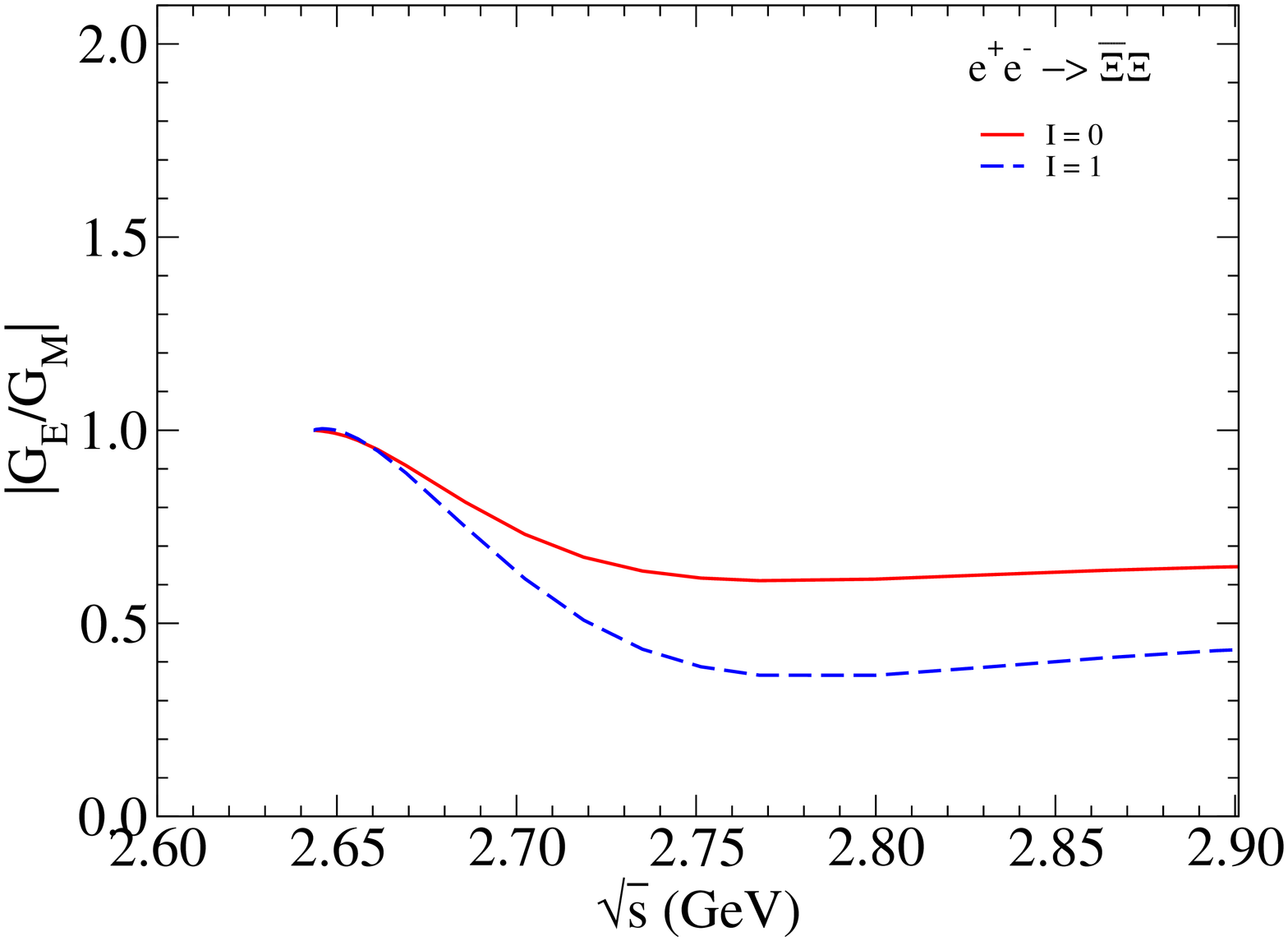}\includegraphics[height=85mm,angle=-90]{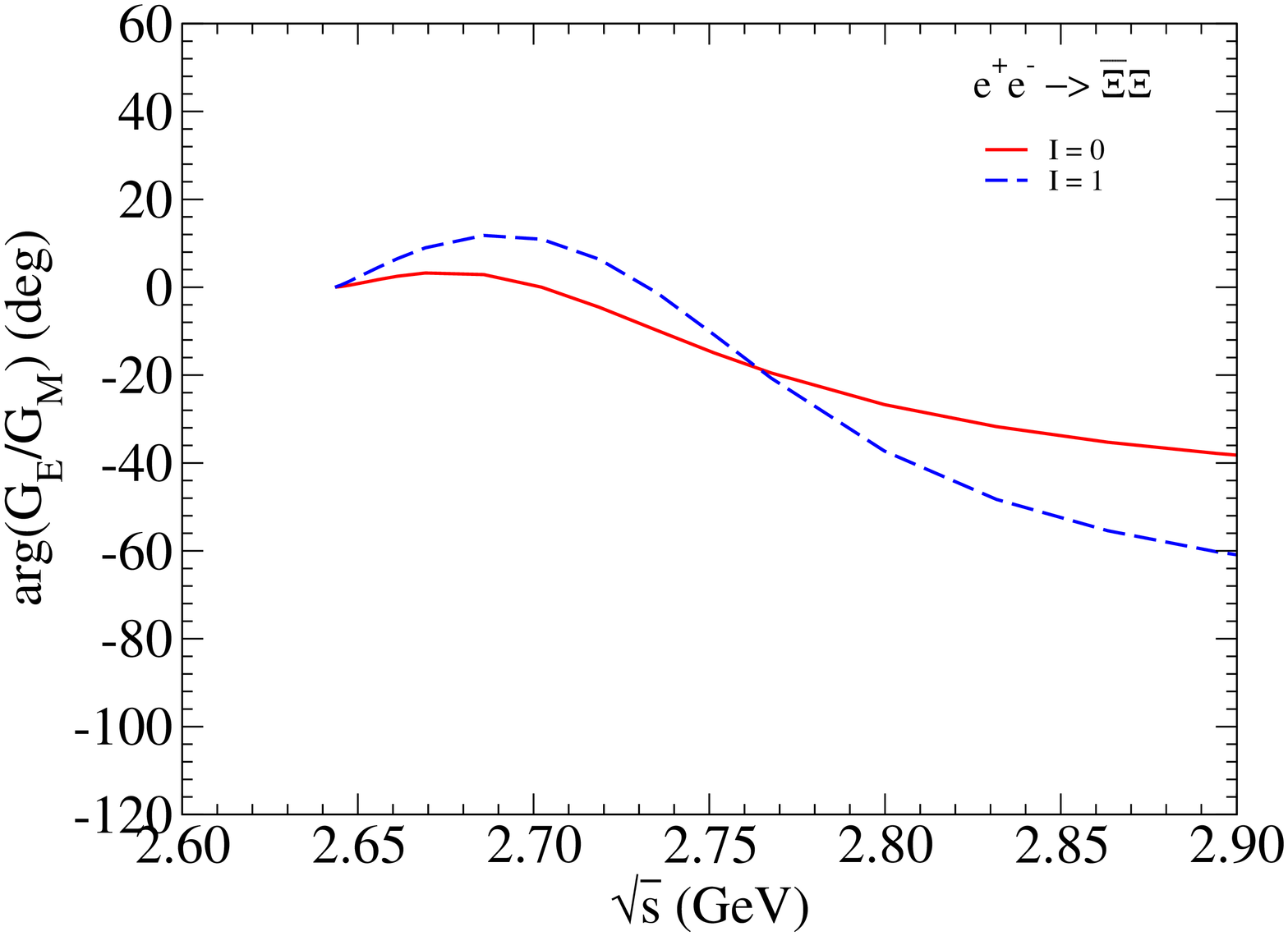}
\caption{$\ebare \to \xbarx$: 
Ratio $|G_E/G_M|$ (left) and phase $\phi=\rm{arg}(G_E/G_M)$ (right).
Same description of curves as in Fig.~\ref{fig:XX}. 
}
\label{fig:RaXX}
\end{center}
\end{figure}

\subsection{{\boldmath$\ebare \to \xbarx$}}

Finally, and as a curiosity, we present here also results for the reaction
$\ebare \to \xbarx$, see Figs.~\ref{fig:XX} and \ref{fig:RaXX}.  
Those are based on a model used in an exploratory calculation for
$\pbarp \to \xbarx$ \cite{Haidenbauer:1993X}. However, since only upper
limits for the cross sections are availabe, it is clear that the 
$\xbarx$ interaction established in that work is basically the result of 
an educated guess and not really constrained by empirical information. 
With only experimental information on $\ebare \to \xmbarxm$ the relative
weight of the $\xmbarxm$ and $\xobarxo$ channels cannot be fixed and, 
therefore, for simplicity we show the results in the isospin channels. 
In order to guide the eye we normalized the curves to the data point at $2.9$~GeV. 
In any case, as one can see in Fig.~\ref{fig:XX}, the predictions for the
cross sections and for $|G_{\rm eff}|$ are fairly similar for the two isospins. 
There is more sensitivity to the details in the ratio $|G_E/G_M|$ and phase 
$\phi=\rm{arg}(G_E/G_M)$, see Figs.~\ref{fig:RaXX}. 
A comparison with the cross sections from BESIII suggests that the FSI effects 
generated from the employed $\xbarx$ model are, may be, somewhat too strong.
Obviously additional data points, say in the region $2.65$-$2.75$~GeV, would be 
very helpful for drawing reliable conclusions.   

\section{Summary}

In the present paper we studied the electromagnetic form factors 
of hyperons in the timelike region, accessible in the reaction $\ebare \to \ybary$.
We focussed on energies near the reaction thresholds and put specific 
emphasis on the role played by the interaction in the final $\ybary$ state. 
The calculation is based on the one-photon approximation for the elementary
reaction mechanism, but takes into account rigorously the effects of 
the interactions in the $\ybary$ systems, in close analogy to our 
work on $\ebare \to \pbarp$ \cite{Haidenbauer:2014}.
For the $\ybary$ interaction we utilized a variety of potential models
\cite{Haidenbauer:1991,Haidenbauer:1992,Haidenbauer:1993,Haidenbauer:1993X}  
that were established in the analysis of data on the reaction $\pbarp \to \ybary$
provided by the PS185 experiment at LEAR \cite{PS185}. 
Given the wealth of near-threshold data on $\pbarp \to \lbarl$,
we expect that the interaction in the $\lbarl$ channel is fairly well 
constrained.
The situation is much less satisfactory for other $\ybary$ channels like
$\sbarl$ and/or $\sbars$ and, in particular, for $\xbarx$ where the 
$\pbarp \to \ybary$ data base is rather poor. 

Results for the energy dependence of the reaction cross sections and the
effective form factors have been presented. Predictions are provided 
for the electromagnetic form factors $G_M$ and $G_E$ in the timelike
region for the $\Lambda$, $\Sigma$, and $\Xi$ hyperons.
It is found that the energy dependence of the near-threshold $\ebare \to \lbarl$ 
and $\ebare \to \sbarl$ cross sections reported by the BaBar 
collaboration \cite{Aubert:2007} is well reproduced when employing 
various $\lbarl$ and $\sbarl$ potentials from 
\cite{Haidenbauer:1991,Haidenbauer:1993} as final-state interaction. 
Both cross sections exhibit a sharp rise from the threshold and then 
remain basically constant for the next $100$~MeV or so, 
a feature that has been observed also in case of the reactions
$\ebare \to \pbarp$ \cite{Aubert:2006,Lees:2013,Haidenbauer:2014} 
and $\ebare \to \nbarn$ \cite{Achasov:2014,Haidenbauer:2015}. 
This behavior of the cross sections implies an enhancement of the 
effective form factor for energies close to the threshold, 
a property which is likewise reproduced by the employed FSI. 
In case of $\ebare\to\lbarl$ even more delicate observables like the 
ratio $|G_E/G_M|$ and the relative phase between $G_E$ and $G_M$
can be described -- at least at the energy of $2.396$~GeV where the, 
so far, only experimental information is available. 

In view of the fact that there is only rather limited
information on the $\pbarp\to\sbars$ reaction, we have to acknowledge
that the interaction utilized for the $\sbars$ channel is much
less reliable. Thus, the pertinent results reported in the 
present work have primarily an exploratory character. 
Nonetheless, it turned out that the energy dependence of the 
three channels $\ebare \to \spbarsp, \ \sobarso, \ \smbarsm$
at low energies can be roughly reproduced. 
Our calculation suggests that there is a strong interplay between 
the $\spbarsp$, $\sobarso$, and $\smbarsm$ channels  
in the near-threshold region, caused by the $\sbars$ interaction.
Therefore, it would be very interesting to perform further experiments 
that establish reliably the energy dependence 
of the cross sections for all three channels at low energies, but also 
those of other observables like the ratio $|G_E/G_M|$.
The latter seems to be very sensitive to the details of the $\ybary$
interaction and, thus, more extended measurements in the region of the 
$\sbars$ thresholds would definitely provide a deeper insight into 
the $\ybary$ dynamics. 

Finally, let us mention that an alternative and independent source of
information on the $\ybary$ interaction is provided by studies of 
two-particle momentum correlation functions in heavy ion collisions 
and/or high-energy $pp$ collisions. Pertinent
data on $\lbarl$ correlations have been already published by the 
ALICE Collaboration \cite{Acharya:2020}.
Extending those studies to further strange baryon-antibaryon interactions 
such as $\sbars$ and/or $\xbarx$ should be feasible. 
In addition there are 
plans by the ${\rm\overline P}$ANDA Collaboration at FAIR to resume the
measurements of $\pbarp\to \lbarl$, $\pbarp\to \sbarl$, $\pbarp\to \xbarx$
\cite{Schonning:2020,PANDA:2020} that stopped when the LEAR facility
at CERN was shut down.  
Further options to study the $\ybary$ interaction as offered for
example by reactions like $B\to \lbarl K$ \cite{Chang:2009} or 
$B\to \lbarl D$ \cite{Lees:2014} could be also exploited eventually.

\acknowledgements{
This work is supported in part by the Deutsche Forschungsgemeinschaft
(DFG) and the National Natural Science Foundation of China (NSFC) 
through funds provided to the Sino-German CRC 110 ``Symmetries and
the Emergence of Structure in QCD''. The work of UGM was also 
supported by the Chinese Academy of Sciences (CAS) President's
International Fellowship Initiative (PIFI) (Grant No. 2018DM0034)
and by the EU (Strong2020).
L.-Y. Dai is supported by the NSFC with Grant No.~11805059,  
the Joint Large Scale Scientific Facility Funds of the NSFC and CAS 
under Contract No.~U1932110, and Fundamental Research Funds for the 
Central Universities. 
}
 

 \end{document}